\UseRawInputEncoding
\documentclass[prd,aps,nofootinbib,floatfix,11pt]{revtex4}
\usepackage{amsmath,graphicx,epsfig,amssymb,dsfont,mathtools,}
\usepackage[usenames]{color}
\usepackage{ulem} %% for strike-through
\usepackage{bigstrut}
\usepackage{slashed}
\usepackage{multirow}
\usepackage{subfigure}

\allowdisplaybreaks

\newcommand{\tabincell}[2]{\begin{tabular}{@{}#1@{}}#2\end{tabular}}
\begin{document}\title{The Production of Singly Charmed Pentaquark $\bar c q qqq$ from Bottom Baryon}
\author{
   Ye Xing~$^{1}$~\footnote{Email:xingye\_guang@cumt.edu.cn}, Wan-Liang Liu~$^{1}$~\footnote{Email:ts21180010a31@cumt.edu.cn},Yi-Hua Xiao~$^{1}$~\footnote{Email:10203607@cumt.edu.cn}}

\affiliation{$^{1}$  School of Physics, China University of Mining and Technology, Xuzhou 221000, China\\
  }

\begin{abstract}
 In the paper, we study the production of pentaquark $\bar{c}q qqq$ from singly bottom baryon. The tensor representations of pentaquark $\bar{c}q qqq$ are completed at first. Under the light quark symmetry analysis, we decompose the singly charmed pentaquarks into $\bar6$, $15$ and $15'$ multiple states, and deduce the representations of these states. Then we construct the production Hamiltonian of the multiple states in proper order. For completeness, we systematically study the production channels of pentaquark states, including the possible production amplitudes and cross section relations between different channels. Ultimately, we screen out some advantageous channels, which expected to be fairly valuable supports for future search experiments.
\end{abstract}
\maketitle

\section{Introduction}

Recently, a series of pentaquark state candidates were observed from the measurements of LHC experiment, the pentaquark states $P_c(4380)^+$~\cite{LHCb:2015yax}, $P_c(4440)^+$~\cite{LHCb:2019kea}, $P_c(4457)^+$~\cite{LHCb:2019kea} and $P_c(4312)^+$~\cite{LHCb:2019kea} from the analysis of $\Lambda_b^0\to J/\psi pK^-$, the  probable strange pentaquark state $P_{cs}(4459)^0$~\cite{LHCb:2020jpq} in the decay of $\Xi_b^- \to J/\psi \Lambda K^-$ channel, and pentaquark candidate $P_{c}(4337)^+$~\cite{LHCb:2021chn} in an amplitude analysis of $B_s^0 \to J/\psi p \bar p$. Proposed interpretation of the pentaquark states were anticipated in some theoretical models, including the compact pentaquark scenarios~\cite{Santopinto:2016pkp,Deng:2016rus,Ali:2019clg,Ali:2019npk,Maiani:2018tfe}, the molecular models~\cite{Du:2019pij,Wang:2019ato,Chen:2015loa}, the hadro-charmonium model~\cite{Eides:2019tgv}, and kinematical effects~\cite{Guo:2015umn}.
The pentaquark states, consisting of four quarks and an anti-quark, have been predicted by the quark model. Theoretically, exactly as the discovered hidden-charmed pentaquark $\bar c c qqq$$(q=u,d,s)$ above, open-charmed ones $c\bar{q}qqq/\bar c qqqq$ should be possible as well. However, no evidence of them has been found in the current experiments. This draws our interest and precipitates us to inquire about the story of singly charmed pentaquark.
%Systematically study on the nature and structure about the pentaquark can improve the knowledge of hadron physics.
First of all, under the quark model and the "flavor antisymmetry principle", Lipkin suggested that the states $\bar Q suud$ and $\bar Q sudd$ $(Q=c,b)$ are stable~\cite{Lipkin:1987sk}. Then Jaffe and Wilczek estimated the masses of $\bar c ud ud$ and $\bar b ud ud$~\cite{Jaffe:2003sg}. Further more, Wise and collaborators predicted that the bound state $\bar b qq qq/\bar c qq qq$ are stable against strong decays~\cite{Stewart:2004pd}. In addition, the possible bound states were discussed respectively, by one pion exchange interaction~\cite{Yamaguchi:2011xb}, the modified chromo-magnetic interaction model~\cite{An:2020vku}, and the constituent model~\cite{Richard:2019fms}. Until now, the possible existence, mass spectrums and possible decay behaviors of the singly charmed pentaquark have been discussed on many occasions.

Although the theoretical studies devoted to the singly charmed pentaquark have achieved remarkable results, there is no conclusive consensus on the properties, e.g, the masses and stability were determined by differences~\cite{Lipkin:1987sk,Yamaguchi:2011xb,An:2020vku,Richard:2019fms}. In the present work, we will not discuss the differences, instead, the production processes of the singly charmed pentaquark $\bar c qqqq$ with less controversy are the attentions. In principle, the light quark flavor SU(3) symmetry, a model independent method, has been successfully used to meson and baryon system~\cite{Savage:1989ub,Gronau:1995hm,He:1998rq,Chiang:2004nm,Li:2007bh,Wang:2009azc,Cheng:2011qh,Hsiao:2015iiu,Lu:2016ogy,He:2016xvd,Wang:2017vnc,Wang:2017azm,Shi:2017dto,He:2018php,Shi:2020gfp,Li:2021rfj}, consequently to be a convincing tool dealing with $\bar c qqqq$.
Though the SU(3) breaking effects might be sizable, the results can still behave well comparing with the experimental data in a global viewpoint~\cite{Lu:2016ogy,He:2021qnc}. Referring to the LHC experiment, we consider the initial baryon with $bqq$ components, which can form multiple states $\bar3$ and $6$ in SU(3) symmetry. Therefore, the singly charmed pentaquark $\bar c qqqq$, grouped into $\bar 6$, $15$ and $15'$ multiple states, can be produced by b-quark weak decay in initial baryon. Accordingly, we could make a systematical investigation on the pentaquark $\bar c qqqq$, constructing the possible Hamiltonian of production, and screening out several advantageous production channels for the examination in experiment.

The rest of the paper is organized as follows. In Sec.II, we deduce the tensor representations of multiple pentaquark states. We construct the possible production Hamiltonian of multiple pentaquark states in Section III, which including the production from singly bottom baryon $T_{b\bar3}$ or $T_{b6}$ in initial state, for definiteness, we still present a collection of the golden channels. We make a short summary in the end.

\section{Representations of Pentaquark $\bar{c}q qqq$}
\begin{figure}
\includegraphics[width=0.95\columnwidth]{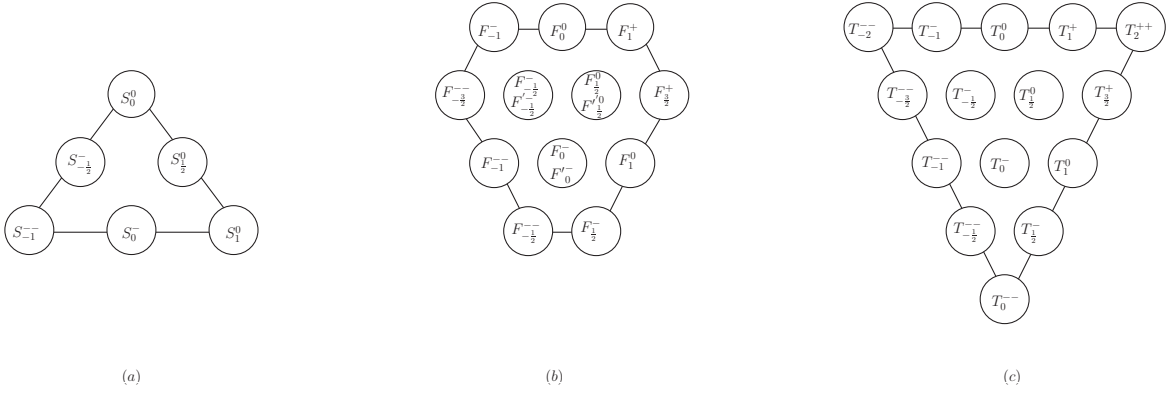}
\caption{The weight graphs of singly charmed pentaquark $\bar c qqqq$, with anti-sextet $\bar 6$ labeled as $S$ in (a), pentagonal states $15$ and $15'$ labeled as $F$ and $T$ in (b) and (c). We recognize them with the electric charge in  upper index and isospin $I_3$ in lower index.}
\label{fig:states}
\end{figure}

Light quark SU(3) flavor symmetry is a good symmetry, especially at the level of hadrons. Aided by the language of group theory, the hadrons can be classified into different group representations, respectively matching with individual spin or orbital quantum number. The singly charmed pentaquark with four light quark $\bar c qqqq$ can then be transformed under the SU(3) symmetry.
\begin{eqnarray}
3 \otimes 3 \otimes 3 \otimes 3=3\oplus \bar{3} \oplus \bar{3} \oplus \bar{6} \oplus \bar{6} \oplus 15 \oplus 15 \oplus 15 \oplus 15'.
\end{eqnarray}
The irreducible representations of new combination states $\bar6$, $15$ and $15'$, arising from group decomposition above, can be shown clearly with tensor reduction, expressing as different tensor forms.
We deduce the tensor decomposition of pentaquark labeled with $T^{ijkl}$,
\begin{eqnarray}\label{eq:tensor}
T^{ijkl}&=& (\hat{T}_3)^{m} (A_1)^{ijkl}_m +(\hat{T}_{\bar3})^m (A_2)^{ijkl}_m +(\hat{T}_{\bar3})^m (A_3)^{ijkl}_m
+ (\overline{T}_{\bar6})_{\{\alpha\beta\}} (B_1)^{ijkl\alpha\beta} +(\overline{T}_{\bar6})_{\{\alpha\beta\}} (B_2)^{ijkl\alpha\beta}\nonumber\\
&&+(\widetilde{T}_{15})_{m}^{\{kl\}} (C_1)^{ijm}+(\widetilde{T}_{15})_{m}^{\{ij\}} (C_2)^{klm}+(\tilde{T}_{15})_m^{\{\alpha\beta\}} (C_3)^{ijklm}_{\alpha\beta}+(T_{15'})^{\{ijkl\}},
\end{eqnarray}
the coefficients of irreducible representations:
\begin{eqnarray*}
&&(A_1)^{ijkl}_m=\frac{1}{8}(\varepsilon^{kli}\delta^j_m-\varepsilon^{klj}\delta^i_m),\
(A_2)^{ijkl}_m=\frac{1}{8}(\varepsilon^{ijk}\delta^l_m+\varepsilon^{ijl}\delta_m^k),\
(A_3)^{ijkl}_m=\frac{1}{8}(\varepsilon^{kli}\delta^j_m+\varepsilon^{klj}\delta_m^i),\\
&&(B_1)^{ijkl\alpha\beta}=\frac{1}{4} \varepsilon^{ij\alpha} \varepsilon^{kl\beta},\
(B_2)^{ijkl\alpha\beta}=\frac{1}{12}\Big(\varepsilon^{lk\alpha}\varepsilon^{ij\beta}+\varepsilon^{ij\alpha}\varepsilon^{lk\beta}-4\varepsilon^{lj\alpha}\varepsilon^{ik\beta}\Big),\\
&&(C_1)^{ijm}=\frac{1}{2}\varepsilon^{ijm},\
(C_2)^{klm}=\frac{1}{2}\varepsilon^{klm},\
(C_3)^{ijklm}_{\alpha\beta}=\frac{1}{6}\Big( 2\delta_{\alpha}^l \delta_{\beta}^j\varepsilon^{ikm}+\delta_{\alpha}^i \delta_{\beta}^j\varepsilon^{lkm}+\delta_{\alpha}^l\delta_{\beta}^k\varepsilon^{ijm}-4\delta_{\alpha}^i\delta_{\beta}^k\varepsilon^{ljm}\Big),
\end{eqnarray*}
here, $\overline{T}_{\bar6}$, $\widetilde{T}_{15}$ and $T_{15'}$ are the irreducible representations of new combination states $\bar 6$, $15$ and $15¡ä$, anti-symmetry index identified as $[ij]$, and symmetry indexes signed with $\{ij\}$. All coefficients are expressed by antisymmetric tensor $\varepsilon$ and tensor $\delta$, both of which are SU(3) invariant symbols.
More specifically, we deduce the possible representations of $\bar 6$,  $15$ and $15'$ states, and list the nonzero components in Tab.~\ref{tab:bar6}. The quark components of the states in flavor space can be reached by expanding the tensor representations, see Appendix~\ref{sec:app}. In addition, we draw the weight graphs of possible states $\bar 6$,  $15$ and $15'$ in Fig.~\ref{fig:states}.
%%%%%%%%%%%%%%%%%%%%%
%%%%%%%%%%%%%%%%%%%%
\begin{table}
\caption{The possible representations of  pentaquark $T_{\bar c q qqq}$ with states $\bar 6$, $15$ and $15'$. Let $S$, $F$ and $T$ be the names of the states respectively. Meanwhile, we give the tensor representations $T_{\bar 6}/T_{15}/T_{15'}$, isospin $I_3$ and hyper-charge Y of the corresponding states.}\label{tab:bar6}
       \begin{tabular}{|c |c c c c|c c c c| c c c c|c|c}\hline\hline
    States& Name & Tensor& $I_3$ & $Y$&Name & Tensor& $I_3$ & $Y$&Name & Tensor& $I_3$ & $Y$ \\\hline
%\multicolumn{12}{|c|} {6 states}\\\hline
\multirow{2}{*}{6 state}&  $S_{0}^{0}$& $T_{\{33\}}$ & 0&0&
    $S_{\frac{1}{2}}^{0}$& $T_{\{23\}}$ & $\frac{1}{2}$&-1&
    $S_{-\frac{1}{2}}^{-}$& $T_{\{13\}}$ & $-\frac{1}{2}$&-1\\
    &$S_{1}^{0}$& $T_{\{22\}}$ & 1&-2&
    $S_{0}^{-}$& $T_{\{12\}}$ & 0&-2&
    $S_{-1}^{--}$& $T_{\{11\}}$ & -1&-2
    \\\hline
%    \multicolumn{12}{|c|} {15 states}\\\hline
\multirow{4}{*}{15 state}&  $F_{1}^{+}$& $T_3^{\{11\}}$ & 1&0&
    $F_{0}^{0}$& $T_3^{\{12\}}$ & 0&0&
    $F_{-1}^{-}$& $T_3^{\{22\}}$ & -1&0\\&
    $F_{\frac{3}{2}}^{+}$& $T_2^{\{11\}}$ & $\frac{3}{2}$&-1&
    $F_{\frac{1}{2}}^{0},{F'}_{\frac{1}{2}}^{0}$& $T_2^{\{12\}},T_3^{\{13\}}$ & $\frac{1}{2}$&-1&
    $F_{-\frac{1}{2}}^{-},{F'}_{-\frac{1}{2}}^{-}$& $T_1^{\{12\}},T_3^{\{23\}}$ & $-\frac{1}{2}$&-1\\&
    $F_{-\frac{3}{2}}^{--}$& $T_1^{\{22\}}$ & $-\frac{3}{2}$&-1&
    $F_{1}^{0}$& $T_2^{\{13\}}$ & $1$&-2&
    $F_{0}^{-},{F'}_{0}^{-}$& $T_1^{\{13\}},T_2^{\{23\}}$ & $0$&-2\\&
    $F_{-1}^{--}$& $T_1^{\{23\}}$ & $-1$&-2&
    $F_{\frac{1}{2}}^{-}$& $T_2^{\{33\}}$ & $\frac{1}{2}$&-3&
    $F_{-\frac{1}{2}}^{--}$& $T_1^{\{33\}}$ & $-\frac{1}{2}$&-3
    \\\hline
%  \multicolumn{12}{|c|} {15' states}\\\hline
\multirow{5}{*}{15' state}&$T_{2}^{++}$& $T^{\{1111\}}$ & 2&0 &$T_{1}^{+}$& $T^{\{1112\}}$ & 1&0 &
                           $T_{0}^{0}$& $T^{\{1122\}}$ & 0&0\\
                           &$T_{-1}^{-}$& $T^{\{1222\}}$ & -1&0&
                           $T_{-2}^{--}$& $T^{\{2222\}}$ & -2&0&$T_{\frac{3}{2}}^{+}$& $T^{\{1113\}}$ & $\frac{3}{2}$&-1\\
                           &$T_{\frac{1}{2}}^{0}$& $T^{\{1123\}}$ &$\frac{1}{2}$&-1 &$T_{-\frac{1}{2}}^{-}$& $T^{\{1223\}}$ & $-\frac{1}{2}$&-1&$T_{-\frac{3}{2}}^{--}$& $T^{\{2223\}}$ & $-\frac{3}{2}$&-1\\
                           &$T_{1}^{0}$& $T^{\{1133\}}$ & 1&-2&$T_{0}^{-}$& $T^{\{1233\}}$ & 0&-2
                           &$T_{-1}^{--}$& $T^{\{2233\}}$ & -1&-2\\
                           &$T_{\frac{1}{2}}^{-}$& $T^{\{1333\}}$ & $\frac{1}{2}$&-3
                           &$T_{-\frac{1}{2}}^{--}$& $T^{\{2333\}}$ & $-\frac{1}{2}$&-3
                           &$T_{0}^{--}$& $T^{\{3333\}}$ & 0&-4\\\hline
%$T_{-\frac{3}{2}}^{--}$& $T^{\{2223\}}$ & $-\frac{3}{2}$&-1\\
%$T_{1}^{0}$& $T^{\{1133\}}$ & 1&-2\\
%$T_{0}^{-}$& $T^{\{1233\}}$ & 0&-2\\
%$T_{-1}^{--}$& $T^{\{2233\}}$ & -1&-2\\
%$T_{\frac{1}{2}}^{-}$& $T^{\{1333\}}$ & $\frac{1}{2}$&-3\\
%$T_{-\frac{1}{2}}^{--}$& $T^{\{2333\}}$ & $-\frac{1}{2}$&-3\\
%$T_{0}^{--}$& $T^{\{3333\}}$ & 0&-4
\end{tabular}
\end{table}

%%%%%%%%%%%%%%%%%%%%%%%%%%%%%%%%%%%%%%
\section{Hamiltonian of productions}
%%%%%%%%%%%%%%%%%%%%%%%%%%%%%%%%%%%%%%
The strategy of SU(3) symmetry analysis needs more representations of hadrons and transition operators, which are pretty easy to achieve. For the singly bottom baryon in initial state, the representations can be decomposed into $\bar 3$ and $6$, found as~\cite{Xing:2019wil,Xing:2021yid}
\begin{eqnarray}
 (T_{\bf{b \bar3}})^{[ij]}= \left(\begin{array}{ccc} 0 & \Lambda_{b}^0  &  \Xi_{b}^0  \\ -\Lambda_{b}^0 & 0 & \Xi_{b}^- \\ -\Xi_{b}^0   &  -\Xi_{b}^-  & 0
  \end{array} \right), \;\;
 (T_{\bf{b6}})^{\{ij\}} = \left(\begin{array}{ccc} \Sigma_{b}^{+} &  \frac{1}{\sqrt{2}}\Sigma_{b}^0   & \frac{1}{\sqrt{2}} \Xi_{b}^{\prime0}\\
  \frac{1}{\sqrt{2}}\Sigma_{b}^0&  \Sigma_{b}^{-} & \frac{1}{\sqrt{2}} \Xi_{b}^{\prime-} \\
  \frac{1}{\sqrt{2}} \Xi_{b}^{\prime0}   &  \frac{1}{\sqrt{2}}  \Xi_{b}^{\prime-}  &  \Omega_{b}^-
  \end{array} \right)\,.
\end{eqnarray}
In the meson sector, singly charmed mesons form an SU(3) triplet or anti-triplet, light mesons form an octet plus singlet, all multiplets are collected as~\cite{Xing:2018bqt,Shi:2020gfp}
\begin{eqnarray}
 M_{8}=\begin{pmatrix}
 \frac{\pi^0}{\sqrt{2}}+\frac{\eta}{\sqrt{6}}
 &\pi^+ & K^+\\
 \pi^-&-\frac{\pi^0}{\sqrt{2}}+\frac{\eta}{\sqrt{6}}&{K^0}\\
 K^-&\bar K^0 &-2\frac{\eta}{\sqrt{6}}
 \end{pmatrix},\;\;
D_i^T = \left(\begin{array}{c}  D^0  \\  D^+  \\  D^+_s
\end{array}\right)\,,\;\;
\overline D^i= \left(\begin{array}{c} \overline D^0  \\  D^-  \\  D^-_s
\end{array}\right).
\end{eqnarray}
The productions in quark level are the transition of $b\to c \bar c d/s$ and $b\to u \bar c d/s$, which can be decomposed into the operators $H_{\bar3}$ and $H_6$, as $(H_{\bar3})_2=V_{cs}^*, (H_6)^{\{13\}}=V_{cs}^*$ for the transition of $b\to u \bar c d/s$, $(H_3)^2=V_{cd}^*, (H_3)^3=V_{cs}^*$  for the transition of $b\to c \bar c d/s$~\cite{He:2016xvd}.
%%%%%%%%%%%%%%%%%%%%%
\begin{figure}
\includegraphics[width=1.00\columnwidth,height=0.18\textheight]{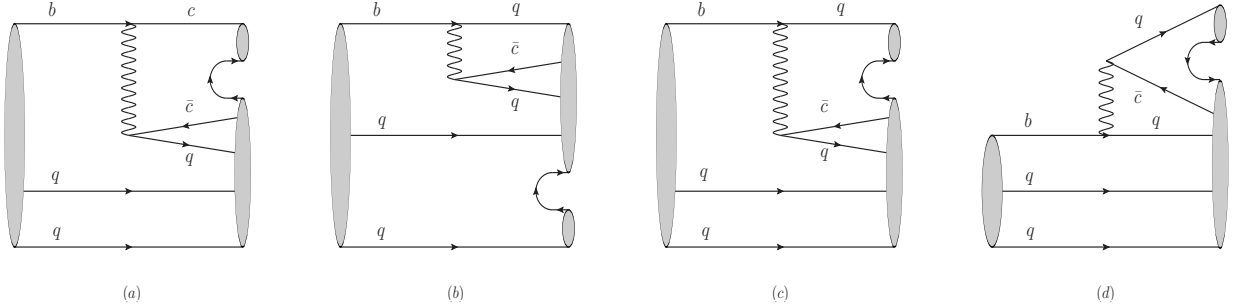}
\caption{The production Feynman diagrams of singly charmed pentaquark multiple states $\bar6$, $15$ and $15'$, starting from singly bottom baryon $bqq$. Diagram(a) represents the process of producing one pentaquark multiple state and a charmed meson. Diagrams(b-d) show the transitions with a pentaquark multiple state and a light meson in final states. }
\label{fig:production}
\end{figure}
%%%%%%%%%%%%%%%%%%%%
\subsection{$\bar6$ states}
Since the representations of hadrons and transition operators have been determined,
we can directly construct the possible Hamiltonian for the production of pentaquark $\bar6$, $15$ and $15'$ states, from the singly bottom baryon, on the hadronic level under the SU(3) symmetry frame. The initial baryon which form multiple state $T_{b\Bar{3}}$ or $T_{b6}$ can produce pentaquark anti-sextet $\Bar{P}_6$ and meson. Accordingly, we construct the possible Hamiltonian for the production of anti-sextet $\Bar{P}_6$, given as follows.
\begin{eqnarray}	\mathcal{H}&&=a_{1}(T_{b\Bar{3}})_{i}(H_{3})^{j}(\Bar{P}_{6})^{\{ik\}}\Bar{D}_{[jk]}+\Bar{a}_{1}(T_{b6})^{\{ij\}}(H_{3})_{[ik]}(\Bar{P}_{6})^{\{kl\}}\Bar{D}_{[jl]}\nonumber\\
&&+d_{1}(T_{b\Bar{3}})_{i}(H_{\Bar{3}})_{j}(\Bar{P}_{6})^{\{ik\}}M_{k}^{j}+d_{2}(T_{b\Bar{3}})_{i}(H_{\Bar{3}})_{j}(\Bar{P}_{6})^{\{jk\}}M_{k}^{i}\nonumber\\ &&+d_{3}(T_{b\Bar{3}})_{i}(H_{6})^{\{ik\}}(\Bar{P}_{6})^{\{jl\}}M_{j}^{m}\varepsilon_{klm}+d_{4}(T_{b\Bar{3}})_{i}(H_{6})^{\{jk\}}(\Bar{P}_{6})^{\{il\}}M_{j}^{m}\varepsilon_{klm}\nonumber\\
&&+\Bar{d}_{1}(T_{b6})^{\{ij\}}(H_{\Bar{3}})_{i}(\Bar{P}_{6})^{\{kl\}}M_{k}^{m}\varepsilon_{jlm}+\Bar{d}_{2}(T_{b6})^{\{ij\}}(H_{\Bar{3}})_{k}(\Bar{P}_{6})^{\{kl\}}M_{i}^{m}\varepsilon_{jlm}\nonumber\\
&&+\Bar{d}_{3}(T_{b6})^{\{ij\}}(H_{6})^{\{kl\}}(\Bar{P}_{6})^{\{\alpha\beta\}}M_{i}^{m}\varepsilon_{\alpha jk}\varepsilon_{\beta lm}+\Bar{d}_{4}(T_{b6})^{\{ij\}}(H_{6})^{\{kl\}}(\Bar{P}_{6})^{\{\alpha\beta\}}M_{k}^{m}\varepsilon_{\alpha il}\varepsilon_{\beta jm}.
\end{eqnarray}	
Here, the independent parameters, $a_1, \Bar{a}_1, d_1, \ldots$, represent non-perturbative effection, $T_{b\bar3}$ or $T_{b6}$ is the multiple state of initial baryon, $H_3$ or $H_6$ is the transition operator, $\Bar{P}_6$ and $\bar D/M$ are respectively the pentaquark anti-sextet and meson in final states. For completeness, the Feynman diagrams for hadronic level are shown in Fig.~\ref{fig:production}.(a-d). Particularly, the Hamiltonian with charmed meson in final states are corresponding with the Feynman diagram Fig.~\ref{fig:production}.(a), by comparison, the diagrams Fig.~\ref{fig:production}.(b-d) are related to the Hamiltonian with light meson in final states.

We expand the Hamiltonian and harvest the possible amplitude results, aggregating into Tab.~\ref{tab:P6_DM}. Typically, the Cabibbo allowed production channels, may receive the largest contribution. Meanwhile, the charged final light meson possesses high detection efficiency. Consequently, we extract these channels and suggest them as golden channels for the studying in the future experiment, see Tab~\ref{tab:golden}. There is no simply way to estimate the size of  production cross section, but the relations between different cross sections can be further deduced, when we approximately ignore the phase space differences. Our calculations show the relations given as follows.
\begin{small}
\begin{eqnarray*}
&&\Gamma( \Lambda_b^0\to S_0^{0} D^0) =\Gamma( \Xi_b^0\to S_{1/2}^{0} D^0) =\Gamma( \Xi_b^-\to S_{-1/2}^{-} D^0)=\Gamma( \Lambda_b^0\to S_{-1/2}^{-} D^+_s)=\Gamma( \Xi_b^0\to S_{0}^- D^+_s) \\
&&=\Gamma( \Xi_b^-\to S_{-1}^{--} D^+_s)=\Gamma( \Lambda_b^0\to   S_0^{0}   D^0) =\Gamma( \Xi_b^0\to   S_{1/2}^{0}   D^0)=\Gamma( \Xi_b^-\to   S_{-1/2}^{-}   D^0)=\Gamma( \Lambda_b^0\to   S_{-1/2}^{-}   D^+_s) \\
&&=\Gamma( \Xi_b^0\to   S_{0}^-   D^+_s) =\Gamma( \Xi_b^-\to   S_{-1}^{--}   D^+_s),\quad \Gamma( \Xi_b^-\to   S_0^{0}  K^-) =\Gamma( \Xi_b^-\to   S_{1/2}^{0}  \pi^-),\\
&&\Gamma( \Lambda_b^0\to S_{-1/2}^{-} D^+) =\Gamma( \Xi_b^0\to S_1^{0} D^0) =\Gamma( \Xi_b^-\to S_{0}^- D^0)=\Gamma( \Lambda_b^0\to S_{1/2}^{0} D^0)=\Gamma( \Xi_b^0\to S_{0}^- D^+) \\
&&=\Gamma( \Xi_b^-\to S_{-1}^{--} D^+)=\Gamma( \Lambda_b^0\to   S_{-1/2}^{-}   D^+) =\Gamma( \Xi_b^0\to   S_1^{0}   D^0)=\Gamma( \Xi_b^-\to   S_{0}^-   D^0)=\Gamma( \Lambda_b^0\to   S_{1/2}^{0}   D^0) \\
&&=\Gamma( \Xi_b^0\to   S_{0}^-   D^+) =\Gamma( \Xi_b^-\to   S_{-1}^{--}   D^+),\quad \Gamma( \Omega_{b}^{-}\to   S_{-1}^{--}  \pi^+) =
2\Gamma( \Xi_{b}^{\prime-}\to   S_{-1}^{--}  K^+),\\
%%%%
&&
\Gamma( \Sigma_{b}^{+}\to S_0^{0} D^+) =2\Gamma( \Sigma_{b}^{0}\to S_0^{0} D^0) =2\Gamma( \Xi_{b}^{\prime0}\to S_{1/2}^{0} D^0) =2\Gamma( \Xi_{b}^{\prime-}\to S_{-1/2}^{-} D^0) =2\Gamma( \Xi_{b}^{\prime-}\to S_{-1}^{--} D^+_s) \\
&&=\Gamma( \Sigma_{b}^{+}\to S_{1/2}^{0} D^+_s)=\Gamma( \Omega_{b}^{-}\to S_{-1}^{--} D^+) =\Gamma( \Omega_{b}^{-}\to S_{0}^- D^0)=2\Gamma( \Xi_{b}^{\prime0}\to S_{0}^- D^+_s)=\frac{1}{2}\Gamma( \Xi_{b}^{\prime0}\to S_{-1/2}^{-} D^+)\\
&&=2\Gamma( \Sigma_{b}^{0}\to S_{-1/2}^{-} D^+_s), \quad \Gamma( \Xi_b^-\to   S_1^{0}  \pi^-) =\Gamma( \Xi_b^-\to   S_{1/2}^{0}  K^-),\quad
\Gamma( \Lambda_b^0\to   S_{-1/2}^{-}  \pi^+) =2\Gamma( \Lambda_b^0\to   S_{1/2}^{0}  \pi^0),\\
&&
2\Gamma( \Xi_{b}^{\prime-}\to S_{0}^- D^0) =\Gamma( \Sigma_{b}^{-}\to S_{-1/2}^{-} D^0) =\Gamma( \Sigma_{b}^{+}\to S_{1/2}^{0} D^+) =
2\Gamma( \Sigma_{b}^{0}\to S_{1/2}^{0} D^0) =\Gamma( \Sigma_{b}^{-}\to S_{-1}^{--} D^+_s) \\
&&=2\Gamma( \Sigma_{b}^{0}\to S_{-1/2}^{-} D^+)=\Gamma( \Sigma_{b}^{+}\to S_1^{0} D^+_s) =\frac{1}{2}\Gamma( \Sigma_{b}^{0}\to S_{0}^- D^+_s) =2\Gamma( \Xi_{b}^{\prime0}\to S_1^{0} D^0) =2\Gamma( \Xi_{b}^{\prime-}\to S_{-1}^{--} D^+) \\
&&2\Gamma( \Xi_{b}^{\prime0}\to S_{0}^- D^+),\quad \Gamma( \Xi_b^0\to   S_{-1/2}^{-}  \pi^+) =\Gamma( \Xi_b^0\to   S_0^{0}  \overline K^0),\quad
\Gamma( \Xi_b^0\to   S_{0}^-  \pi^+) =2\Gamma( \Xi_b^0\to   S_1^{0}  \pi^0),\\
%%%
&&\Gamma( \Sigma_{b}^{-}\to   S_0^{0}  \pi^-) =\Gamma( \Sigma_{b}^{-}\to   S_{-1/2}^{-}  K^0),\quad
\Gamma( \Sigma_{b}^{-}\to   S_{1/2}^{0}  \pi^-) =2\Gamma( \Xi_{b}^{\prime-}\to   S_1^{0}  \pi^-),\\
&&\Gamma( \Sigma_{b}^{0}\to   S_{-1/2}^{-}  \pi^+) =\Gamma( \Sigma_{b}^{-}\to   S_{-1/2}^{-}  \pi^0),\quad
\Gamma( \Sigma_{b}^{-}\to   S_{0}^-  K^0) =2\Gamma( \Sigma_{b}^{0}\to   S_1^{0}  K^0),\\
&&\Gamma( \Sigma_{b}^{-}\to   S_{-1}^{--}  K^+) =2\Gamma( \Xi_{b}^{\prime-}\to   S_{-1}^{--}  \pi^+),\quad
\Gamma( \Omega_{b}^{-}\to   S_{-1/2}^{-}  \overline K^0) =2\Gamma( \Xi_{b}^{\prime0}\to   S_0^{0}  \overline K^0),\\
&&
\Gamma( \Omega_{b}^{-}\to   S_{1/2}^{0}  K^-) =2\Gamma( \Xi_{b}^{\prime-}\to   S_0^{0}  K^-),\quad
\Gamma( \Omega_{b}^{-}\to   S_1^{0}  K^-) =\Gamma( \Omega_{b}^{-}\to   S_{0}^-  \overline K^0),\\
&&\Gamma( \Xi_b^0\to   S_1^{0}  \eta_q) =\Gamma( \Xi_b^-\to   S_{0}^-  \eta_q),\quad
\Gamma( \Lambda_b^0\to   S_{0}^-  K^+) =\Gamma( \Lambda_b^0\to   S_1^{0}  K^0),\\
&&
\Gamma( \Xi_{b}^{\prime0}\to   S_{0}^-  \pi^+) =2\Gamma( \Xi_{b}^{\prime0}\to   S_1^{0}  \pi^0),\quad
\Gamma( \Xi_{b}^{\prime-}\to   S_{0}^-  \eta_q) =\Gamma( \Xi_{b}^{\prime0}\to   S_1^{0}  \eta_q).
\end{eqnarray*}	
\end{small}

\begin{table}
\tiny
\caption{The productions of pentaquark anti-sextet $P_{\bar6}$ from singly bottom baryons $T_{b\bar3}/T_{b6}$, with charmed meson in the final state.}\label{tab:P6_DM}\begin{tabular}{|ll|ll|ll|}\hline\hline
channel & amplitude &channel &amplitude &channel &amplitude\\\hline
$\Lambda_b^0\to S_0^{0} D^0 $ & $ a_1 V_{cd}^*$&
$\Xi_b^0\to S_{1/2}^{0} D^0 $ & $ -a_1 V_{cd}^*$&
$\Xi_b^-\to S_{-1/2}^{-} D^0 $ & $ a_1 V_{cd}^*$\\

$\Lambda_b^0\to S_{-1/2}^{-} D^+_s/D^+ $ & $ -a_1 V_{cs}^*/a_1 V_{cd}^*$&
$\Xi_b^0\to S_1^{0} D^0 $ & $ a_1 V_{cs}^*$&
$\Xi_b^-\to S_{0}^- D^0 $ & $ -a_1 V_{cs}^*$\\

$\Lambda_b^0\to S_{1/2}^{0} D^0 $ & $ -a_1 V_{cs}^*$&
$\Xi_b^0\to S_{0}^- D^+/D^+_s $ & $ -a_1 V_{cs}^*/a_1 V_{cd}^*$&
$\Xi_b^-\to S_{-1}^{--} D^+_s/D^+ $ & $ -a_1 V_{cd}^*/a_1 V_{cs}^*$\\

$\Lambda_b^0\to   S_{0}^-  K^+  $ & $ (d_3-d_2) V_{cs}^*$&
$\Xi_b^0\to   S_{0}^-  K^+  $ & $ -(d_1+d_3-d_4) V_{cd}^*$&
$\Xi_b^-\to   S_{0}^-  \eta_q  $ & $ -\frac{(d_1+d_2+3 d_4) V_{cs}^*}{\sqrt{6}}$\\
%%%%%%%%%%%%%%%%%%%%%%%%
$\Lambda_b^0\to   S_0^{0}  \pi^0  $ & $ \sqrt{2} d_4 V_{cd}^*$&
$\Xi_b^0\to   S_0^{0}  \overline K^0  $ & $ (d_3-d_2) V_{cd}^*$&
$\Xi_b^-\to   S_0^{0}  K^-  $ & $ (d_2+d_3) V_{cd}^*$\\

$\Lambda_b^0\to   S_0^{0}  \overline K^0  $ & $ -(d_1+d_3+d_4) V_{cs}^*$&
$\Xi_b^0\to   S_{1/2}^{0}  \pi^0  $ & $  \frac{( d_2 -d_3 -2 d_4)  V_{cd}^*}{\sqrt{2}}$&
$\Xi_b^-\to   S_{1/2}^{0}  \pi^-  $ & $ (d_2+d_3) V_{cd}^*$\\

$\Lambda_b^0\to   S_0^{0}  \eta_q  $ & $ -\sqrt{\frac{2}{3}} (d_1+d_2) V_{cd}^*$&
$\Xi_b^0\to   S_{1/2}^{0}  \overline K^0  $ & $ (d_1+d_2+d_4) V_{cs}^*$&
$\Xi_b^-\to   S_{1/2}^{0}  K^-  $ & $ -(d_2+d_3) V_{cs}^*$\\

$\Lambda_b^0\to   S_{1/2}^{0}  \pi^0  $ & $ \frac{(d_1+d_3-d_4) V_{cs}^*}{\sqrt{2}}$&
$\Xi_b^0\to   S_{1/2}^{0}  \eta_q  $ & $ \frac{(2 d_1-d_2+3 d_3) V_{cd}^*}{\sqrt{6}}$&
$\Xi_b^-\to   S_{-1/2}^{-}  \pi^0  $ & $ \frac{(d_2+d_3+2 d_4) V_{cd}^*}{\sqrt{2}}$\\

$\Lambda_b^0\to   S_{1/2}^{0}  K^0  $ & $ (d_1+d_2+d_4) V_{cd}^*$&
$\Xi_b^0\to   S_{-1/2}^{-}  \pi^+  $ & $ (d_3-d_2) V_{cd}^*$&
$\Xi_b^-\to   S_{-1/2}^{-}  \overline K^0  $ & $ (d_3-d_1-d_4) V_{cs}^*$\\

$\Lambda_b^0\to   S_{1/2}^{0}  \eta_q  $ & $ \frac{(2 d_2-d_1-3 (d_3+d_4)) V_{cs}^*}{\sqrt{6}}$&
$\Xi_b^0\to   S_1^{0}  \pi^0  $ & $ \frac{-(d_1+d_2-d_4) V_{cs}^*}{\sqrt{2}}$&
$\Xi_b^-\to   S_{-1/2}^{-}  \eta_q  $ & $ \frac{(-2 d_1+d_2+3 d_3) V_{cd}^*}{\sqrt{6}}$\\

$\Lambda_b^0\to   S_{-1/2}^{-}  \pi^+  $ & $ -(d_1+d_3-d_4) V_{cs}^*$&
$\Xi_b^0\to   S_1^{0}  K^0  $ & $ -(d_1+d_3+d_4) V_{cd}^*$&
$\Xi_b^-\to   S_1^{0}  \pi^-  $ & $ -(d_2+d_3) V_{cs}^*$\\

$\Lambda_b^0\to   S_{-1/2}^{-}  K^+  $ & $ (d_1+d_2-d_4) V_{cd}^*$&
$\Xi_b^0\to   S_1^{0}  \eta_q  $ & $ \frac{(d_1+d_2+3 d_4) V_{cs}^*}{\sqrt{6}}$&
$\Xi_b^-\to   S_{0}^-  \pi^0  $ & $ \frac{(d_1-d_2-2 d_3-d_4) V_{cs}^*}{\sqrt{2}}$\\

$\Lambda_b^0\to   S_1^{0}  K^0  $ & $ (d_3-d_2) V_{cs}^*$&
$\Xi_b^0\to   S_{0}^-  \pi^+  $ & $ (d_1+d_2-d_4) V_{cs}^*$&
$\Xi_b^-\to   S_{0}^-  K^0  $ & $ (d_1-d_3+d_4) V_{cd}^*$\\

&&&&
$\Xi_b^-\to   S_{-1}^{--}  K^+  $ & $ (d_1-d_3-d_4) V_{cd}^*$\\

&&&&
$\Xi_b^-\to   S_{-1}^{--}  \pi^+  $ & $ (-d_1+d_3+d_4) V_{cs}^*$\\
\hline
%%%%%%%%%%%%%%%%%%%%%%%%%%%%%%%%%%%%%%%%%%%%%%%%%%%%%%%%%
$\Sigma_{b}^{+}\to S_0^{0} D^+ $ & $ \bar{a}_1 V_{\text{cd}} {}^*$&
$\Omega_{b}^{-}\to S_{0}^- D^0 $ & $ -\bar{a}_1 V_{\text{cd}} {}^*$&
$\Sigma_{b}^{-}\to S_{-1/2}^{-} D^0 $ & $ -\bar{a}_1 V_{cs}^*$\\

$\Sigma_{b}^{+}\to S_{1/2}^{0} D^+ $ & $ -\bar{a}_1 V_{\text{cs}} {}^*$&
$\Omega_{b}^{-}\to S_{-1}^{--} D^+ $ & $ \bar{a}_1 V_{\text{cd}} {}^*$&
$\Sigma_{b}^{-}\to S_{-1}^{--} D^+_s $ & $ \bar{a}_1 V_{cs}^*$\\

$\Sigma_{b}^{+}\to S_{1/2}^{0} D^+_s $ & $ -\bar{a}_1 V_{\text{cd}} {}^*$&
$\Omega_{b}^{-}\to   S_{1/2}^{0}  K^-  $ & $ -(\bar{d}_1+\bar{d}_2+\bar{d}_3) V_{cd}^*$&
$\Sigma_{b}^{-}\to   S_{-1}^{--}  K^+  $ & $ (\bar{d}_1+\bar{d}_4) V_{cs}^*$\\

$\Sigma_{b}^{+}\to S_1^{0} D^+_s $ & $ \bar{a}_1 V_{cs}^*$&
$\Omega_{b}^{-}\to   S_{-1/2}^{-}  \overline K^0  $ & $ (\bar{d}_1+\bar{d}_2-\bar{d}_3) V_{cd}^*$&
$\Sigma_{b}^{-}\to   S_{0}^-  K^0  $ & $ (\bar{d}_1+\bar{d}_2-\bar{d}_3) V_{cs}^*$\\
%%%%%%%%%%%%%%%%%%
$\Sigma_{b}^{+}\to   S_0^{0}  \pi^+  $ & $ (-\bar{d}_2+\bar{d}_3+\bar{d}_4) V_{cd}^*$&
$\Omega_{b}^{-}\to   S_1^{0}  K^-  $ & $ (\bar{d}_2+\bar{d}_3+\bar{d}_4) V_{cs}^*$&
$\Sigma_{b}^{-}\to   S_0^{0}  \pi^-  $ & $ (\bar{d}_2+\bar{d}_3+\bar{d}_4) V_{cd}^*$\\

$\Sigma_{b}^{+}\to   S_{1/2}^{0}  \pi^+  $ & $ (\bar{d}_2-\bar{d}_3-\bar{d}_4) V_{cs}^*$&
$\Omega_{b}^{-}\to   S_{0}^-  \pi^0  $ & $ -\sqrt{2} \bar{d}_1 V_{cd}^*$&
$\Sigma_{b}^{-}\to   S_0^{0}  K^-  $ & $ (\bar{d}_4-\bar{d}_1) V_{cs}^*$\\

$\Sigma_{b}^{+}\to   S_{1/2}^{0}  K^+  $ & $ (\bar{d}_2-\bar{d}_3-\bar{d}_4) V_{cd}^*$&
$\Omega_{b}^{-}\to   S_{0}^-  \overline K^0  $ & $ -(\bar{d}_2+\bar{d}_3+\bar{d}_4) V_{cs}^*$&
$\Sigma_{b}^{-}\to   S_{1/2}^{0}  \pi^-  $ & $ -(\bar{d}_1+\bar{d}_2+\bar{d}_3) V_{cs}^*$\\

$\Sigma_{b}^{+}\to   S_1^{0}  K^+  $ & $ (-\bar{d}_2+\bar{d}_3+\bar{d}_4) V_{cs}^*$&
$\Omega_{b}^{-}\to   S_{0}^-  \eta_q  $ & $ -\sqrt{\frac{2}{3}} (2 \bar{d}_3+\bar{d}_4) V_{cd}^*$&
$\Sigma_{b}^{-}\to   S_{-1/2}^{-}  \pi^0  $ & $ -\frac{(\bar{d}_1+2 \bar{d}_3+\bar{d}_4) V_{cs}^*}{\sqrt{2}}$\\

&&
$\Omega_{b}^{-}\to   S_{-1}^{--}  \pi^+  $ & $ (\bar{d}_1+\bar{d}_4) V_{cd}^*$&
$\Sigma_{b}^{-}\to   S_{-1/2}^{-}  K^0  $ & $ -(\bar{d}_2+\bar{d}_3+\bar{d}_4) V_{cd}^*$\\

&&
$\Omega_{b}^{-}\to   S_1^{0}  \pi^-  $ & $ (\bar{d}_4-\bar{d}_1) V_{cd}^*$&
$\Sigma_{b}^{-}\to   S_{-1/2}^{-}  \eta_q  $ & $ \frac{(-3 \bar{d}_1+2 \bar{d}_3+\bar{d}_4) V_{cs}^*}{\sqrt{6}}$\\\hline
%%%%%%%%%%%%%%%%%%%%%%%%%%%%%%%%%%%%%%%%%%%%%%%%%%%%%%
%%%%%%%%%%%%%%%%%%%%%%%%%%%%%%%%%%%%%%%%%%%%%%%%%%%%%%%%%%
$\Xi_{b}^{\prime0}\to S_{1/2}^{0} D^0 $ & $ \frac{\bar{a}_1 V_{\text{cd}} {}^*}{\sqrt{2}}$&
$\Xi_{b}^{\prime-}\to S_{-1/2}^{-} D^0 $ & $ \frac{\bar{a}_1 V_{\text{cd}} {}^*}{\sqrt{2}}$&
$\Sigma_{b}^{0}\to S_0^{0} D^0 $ & $ -\frac{\bar{a}_1 V_{\text{cd}} {}^*}{\sqrt{2}}$\\

$\Xi_{b}^{\prime0}\to S_{-1/2}^{-} D^+ $ & $ -\sqrt{2} \bar{a}_1 V_{\text{cd}} {}^*$&
$\Xi_{b}^{\prime-}\to S_{0}^- D^0 $ & $ \frac{\bar{a}_1 V_{cs}^*}{\sqrt{2}}$&
$\Sigma_{b}^{0}\to S_{1/2}^{0} D^0 $ & $ \frac{\bar{a}_1 V_{cs}^*}{\sqrt{2}}$\\

$\Xi_{b}^{\prime0}\to S_1^{0} D^0 $ & $ -\frac{\bar{a}_1 V_{cs}^*}{\sqrt{2}}$&
$\Xi_{b}^{\prime-}\to S_{-1}^{--} D^+/D^+_s $ & $ -\frac{\bar{a}_1 V_{cs}^*/\bar{a}_1 V_{cd}^*}{\sqrt{2}}$&
$\Sigma_{b}^{0}\to S_{-1/2}^{-} D^+/D^+_s $ & $ \frac{\bar{a}_1 V_{cs}^*/\bar{a}_1 V_{cd}^*}{\sqrt{2}}$
\\

$\Xi_{b}^{\prime0}\to S_{0}^- D^+/D^+_s $ & $ \frac{\bar{a}_1 V_{cs}^*/\bar{a}_1 V_{cd}^*}{\sqrt{2}}$&
$\Xi_{b}^{\prime-}\to   S_{0}^-  \eta_q  $ & $ -\frac{(3 \bar{d}_2-\bar{d}_3+\bar{d}_4) V_{cs}^*}{2 \sqrt{3}}$&
$\Sigma_{b}^{0}\to   S_{1/2}^{0}  K^0  $ & $ \frac{(\bar{d}_2-\bar{d}_3+\bar{d}_4) V_{cd}^*}{\sqrt{2}}$
\\

$\Xi_{b}^{\prime0}\to   S_0^{0}  \overline K^0  $ & $ -\frac{(\bar{d}_1+\bar{d}_2-\bar{d}_3) V_{cd}^*}{\sqrt{2}}$&
$\Xi_{b}^{\prime-}\to   S_{-1}^{--}  K^+  $ & $ -\frac{(\bar{d}_1+\bar{d}_4) V_{cd}^*}{\sqrt{2}}$
&$\Sigma_{b}^{0}\to S_{0}^- D^+_s $ & $ -\sqrt{2} \bar{a}_1 V_{cs}^*$
\\
%%%%%%%%%%%%%%%%%%%%%%%%%%%%%%%%%%%%%%%%%%%%%%%%%%%%%%%%%%
%%%%%%%%%%%%%%%%%%%%%%%%%%%%%%%%%%%%%%%%%%%%%%%%%%%%%%%%%%%%

$\Xi_{b}^{\prime0}\to   S_{1/2}^{0}  \pi^0  $ & $ \frac{ (\bar{d}_1-\bar{d}_2-\bar{d}_3) V_{cd}^*}{2}$&
$\Xi_{b}^{\prime-}\to   S_0^{0}  K^-  $ & $ \frac{(\bar{d}_1+\bar{d}_2+\bar{d}_3) V_{cd}^*}{\sqrt{2}}$&
$\Sigma_{b}^{0}\to   S_{0}^-  K^+  $ & $ \frac{-(\bar{d}_1-\bar{d}_2+\bar{d}_3+2 \bar{d}_4) V_{cs}^*}{\sqrt{2}}$\\

$\Xi_{b}^{\prime0}\to   S_{1/2}^{0}  \overline K^0  $ & $ \frac{(\bar{d}_2-\bar{d}_3+\bar{d}_4) V_{cs}^*}{\sqrt{2}}$&
$\Xi_{b}^{\prime-}\to   S_{1/2}^{0}  \pi^-  $ & $ \frac{(\bar{d}_1-\bar{d}_2-\bar{d}_3-2 \bar{d}_4) V_{cd}^*}{\sqrt{2}}$&
$\Sigma_{b}^{0}\to   S_1^{0}  K^0  $ & $ -\frac{(\bar{d}_1+\bar{d}_2-\bar{d}_3) V_{cs}^*}{\sqrt{2}}$\\

$\Xi_{b}^{\prime0}\to   S_{1/2}^{0}  \eta_q  $ & $ \frac{(\bar{d}_3-3 \bar{d}_1-3 \bar{d}_2+2 \bar{d}_4) V_{cd}^*}{2 \sqrt{3}}$&
$\Xi_{b}^{\prime-}\to   S_{1/2}^{0}  K^-  $ & $ \frac{(\bar{d}_1-\bar{d}_2-\bar{d}_3-2 \bar{d}_4) V_{cs}^*}{\sqrt{2}}$&
$\Sigma_{b}^{0}\to   S_{-1/2}^{-}  \pi^+  $ & $ \frac{(\bar{d}_1+2 \bar{d}_3+\bar{d}_4) V_{cs}^*}{\sqrt{2}}$\\

$\Xi_{b}^{\prime0}\to   S_{0}^-  K^+  $ & $ \frac{(\bar{d}_1+2 \bar{d}_3+\bar{d}_4) V_{cd}^*}{\sqrt{2}}$&
$\Xi_{b}^{\prime-}\to   S_{-1/2}^{-}  \pi^0  $ & $ \frac{ (\bar{d}_1-\bar{d}_2+\bar{d}_3) V_{cd}^*}{2}$&
$\Sigma_{b}^{0}\to   S_{-1/2}^{-}  K^+  $ & $ -\frac{(\bar{d}_2+\bar{d}_3-\bar{d}_4) V_{cd}^*}{\sqrt{2}}$\\

$\Xi_{b}^{\prime0}\to   S_{-1/2}^{-}  \pi^+  $ & $ \frac{-(\bar{d}_1-\bar{d}_2+\bar{d}_3+2 \bar{d}_4) V_{cd}^*}{\sqrt{2}}$&
$\Xi_{b}^{\prime-}\to   S_{-1/2}^{-}  \overline K^0  $ & $ \frac{(-\bar{d}_1+2 \bar{d}_3+\bar{d}_4) V_{cs}^*}{\sqrt{2}}$&
$\Sigma_{b}^{0}\to   S_{1/2}^{0}  \eta_q  $ & $ \frac{(3 \bar{d}_1-2 \bar{d}_3-\bar{d}_4) V_{cs}^*}{2 \sqrt{3}}$\\

$\Xi_{b}^{\prime0}\to   S_1^{0}  \pi^0  $ & $ \frac{ (\bar{d}_2+\bar{d}_3-\bar{d}_4) V_{cs}^*}{2}$&
$\Xi_{b}^{\prime-}\to   S_{-1/2}^{-}  \eta_q  $ & $ \frac{(3 \bar{d}_1+3 \bar{d}_2+\bar{d}_3+2 \bar{d}_4) V_{cd}^*}{2 \sqrt{3}}$&
$\Sigma_{b}^{0}\to   S_0^{0}  \pi^0  $ & $ \bar{d}_2 V_{cd}^*$\\

$\Xi_{b}^{\prime0}\to   S_1^{0}  K^0  $ & $ \frac{(\bar{d}_1-\bar{d}_4) V_{cd}^*}{\sqrt{2}}$&
$\Xi_{b}^{\prime-}\to   S_1^{0}  \pi^-  $ & $ \frac{(\bar{d}_1+\bar{d}_2+\bar{d}_3) V_{cs}^*}{\sqrt{2}}$&
$\Sigma_{b}^{0}\to   S_0^{0}  \overline K^0  $ & $ \frac{(\bar{d}_1-\bar{d}_4) V_{cs}^*}{\sqrt{2}}$\\

$\Xi_{b}^{\prime0}\to   S_1^{0}  \eta_q  $ & $ \frac{(3 \bar{d}_2-\bar{d}_3+\bar{d}_4) V_{cs}^*}{2 \sqrt{3}}$&
$\Xi_{b}^{\prime-}\to   S_{0}^-  \pi^0  $ & $ \frac{(2 \bar{d}_1+\bar{d}_2+\bar{d}_3+\bar{d}_4) V_{cs}^*}{2}$&
$\Sigma_{b}^{0}\to   S_0^{0}  \eta_q  $ & $ \frac{(\bar{d}_3-\bar{d}_4) V_{cd}^*}{\sqrt{3}}$\\

$\Xi_{b}^{\prime0}\to   S_{0}^-  \pi^+  $ & $ -\frac{(\bar{d}_2+\bar{d}_3-\bar{d}_4) V_{cs}^*}{\sqrt{2}}$&
$\Xi_{b}^{\prime-}\to   S_{0}^-  K^0  $ & $ \frac{(-\bar{d}_1+2 \bar{d}_3+\bar{d}_4) V_{cd}^*}{\sqrt{2}}$&
$\Sigma_{b}^{0}\to   S_{1/2}^{0}  \pi^0  $ & $ \frac{-(\bar{d}_1+2 \bar{d}_2-\bar{d}_4) V_{cs}^*}{2}$\\

&&
$\Xi_{b}^{\prime-}\to   S_{-1}^{--}  \pi^+  $ & $ -\frac{(\bar{d}_1+\bar{d}_4) V_{cs}^*}{\sqrt{2}}$
&&\\
\hline
\end{tabular}
\end{table}

\subsection{15 states}
Similarly, we construct the possible Hamiltonian for the production of pentaquark $15$ state from the singly bottom baryon.
\begin{eqnarray} \mathcal{H}&&=b_{1}(T_{b\Bar{3}})_{i}(H_{3})^{j}(\Bar{P}_{15})_{\{jk\}}^{i}\Bar{D}^{k}+\Bar{b}_{1}(T_{b6})^{\{ij\}}(H_{3})^{k}(\Bar{P}_{15})_{\{ij\}}^{l}\Bar{D}_{[kl]}
+\Bar{b}_{2}(T_{b6})^{\{ij\}}(H_{3})^{k}(\Bar{P}_{15})_{\{ik\}}^{l}\Bar{D}_{[jl]}\nonumber\\ &&+e_{1}(T_{b\Bar{3}})_{i}(H_{\Bar{3}})^{[ij]}(\Bar{P}_{15})_{\{jk\}}^{l}M_{l}^{k}+e_{2}(T_{b\Bar{3}})_{i}(H_{\Bar{3}})^{[jl]}(\Bar{P}_{15})_{\{jk\}}^{i}M_{l}^{k}+e_{3}(T_{b\Bar{3}})_{i}(H_{6})^{\{ij\}}(\Bar{P}_{15})_{\{jk\}}^{l}M_{l}^{k}\nonumber\\
&&+e_{4}(T_{b\Bar{3}})_{i}(H_{6})^{\{jl\}}(\Bar{P}_{15})_{\{jk\}}^{i}M_{l}^{k}+e_{5}(T_{b\Bar{3}})_{i}(H_{6})^{\{jk\}}(\Bar{P}_{15})_{\{jk\}}^{l}M_{l}^{i}+\Bar{e}_{1}(T_{b6})^{\{ij\}}(H_{\Bar{3}})_{i}(\Bar{P}_{15})_{\{jl\}}^{k}M_{k}^{l}\nonumber\\
&&+\Bar{e}_{2}(T_{b6})^{\{ij\}}(H_{\Bar{3}})_{k}(\Bar{P}_{15})_{\{il\}}^{k}M_{j}^{l}+\Bar{e}_{3}(T_{b6})^{\{ij\}}(H_{\Bar{3}})_{l}(\Bar{P}_{15})_{\{ij\}}^{k}M_{k}^{l}+\Bar{e}_{4}(T_{b6})^{\{ij\}}(H_{6})^{\{kl\}}(\Bar{P}_{15})_{\{ij\}}^{\alpha}M_{k}^{m}\varepsilon_{\alpha lm}\nonumber\\
	&&+\Bar{e}_{5}(T_{b6})^{\{ij\}}(H_{6})^{\{kl\}}(\Bar{P}_{15})_{\{ik\}}^{\alpha}M_{j}^{m}\varepsilon_{\alpha lm}+\Bar{e}_{6}(T_{b6})^{\{ij\}}(H_{6})^{\{kl\}}(\Bar{P}_{15})_{\{im\}}^{\alpha}M_{k}^{m}\varepsilon_{\alpha jl}\nonumber\\
	&&+\Bar{e}_{7}(T_{b6})^{\{ij\}}(H_{6})^{\{kl\}}(\Bar{P}_{15})_{\{kl\}}^{\alpha}M_{i}^{m}\varepsilon_{\alpha jm}+\Bar{e}_{8}(T_{b6})^{\{ij\}}(H_{6})^{\{kl\}}(\Bar{P}_{15})_{\{km\}}^{\alpha}M_{i}^{m}\varepsilon_{\alpha jl}.
\end{eqnarray}
The corresponding Feynman diagrams are shown in Fig.~\ref{fig:production}. We collect the possible channels and amplitude results into Tab~\ref{tab:P15_D} and Tab~\ref{tab:P15_M}. Several golden channels for producing pentaquark 15 state are arranged in Tab~\ref{tab:golden}. For convenience, the cross section relations between different channels with anti-triplet or sextet baryon $T_{b\bar3}/T_{b6}$ in initial state are reorganized in Appendix~\ref{sec:table}.

\subsection{15' states}
The possible Hamiltonian for the production of pentaquark $15'$ state on the hadronic level under the SU(3) symmetry frame are constructed as below.
\begin{eqnarray}
\mathcal{H}&&=\Bar{c}_{1}(T_{b6})^{\{ij\}}(H_{3})^{k}(\Bar{P}_{15^{\prime}})_{\{ijkl\}}\Bar{D}^{l}+f_{1}(T_{b\Bar{3}})^{[im]}(H_{\Bar{3}})^{[jl]}(\Bar{P}_{15^{\prime}})_{\{ijkl\}}M_{m}^{k}\nonumber\\
&&+f_{2}(T_{b\Bar{3}})^{[im]}(H_{6})^{\{jk\}}(\Bar{P}_{15^{\prime}})_{\{ijkl\}}M_{m}^{l}+\Bar{f}_{1}(T_{b6})^{\{ij\}}(H_{\Bar{3}})^{[km]}(\Bar{P}_{15^{\prime}})_{\{ijkl\}}M_{m}^{l}\nonumber\\
&&+\Bar{f}_{2}(T_{b6})^{\{ij\}}(H_{6})^{\{km\}}(\Bar{P}_{15^{\prime}})_{\{ijkl\}}M_{m}^{l}+\Bar{f}_{3}(T_{b6})^{\{im\}}(H_{6})^{\{jk\}}(\Bar{P}_{15^{\prime}})_{\{ijkl\}}M_{m}^{l}.
\end{eqnarray}
We draw the Feynman diagrams in Fig.~\ref{fig:production}. It should be noted that one anti-triplet baryon $T_{b\bar3}$ can not produce the pentaquark $15'$ state, because the symmetry and anti-symmetry indexes in the only allowed Hamiltonian $(T_{b\Bar{3}})^{[ij]}(H_{3})^{k}(\Bar{P}_{15^{\prime}})_{\{ijkl\}}\Bar{D}^{l}$ forbid the process. The possible channels and amplitude results are collected into Tab~\ref{tab:P15p_DM}. The golden channels for the pentaquark $15'$ state are fixed in Tab~\ref{tab:golden}. We deduce the cross section relations between different channels, consistently written in Appendix~\ref{sec:table}.

\begin{table}
\caption{The golden channels for producing singly charmed pentaquark multiple states $\bar6$, $15$ and $15'$, with the charmed meson or light meson in final states.}\label{tab:golden}
\begin{tabular}{|c c c c c| c c|}\hline\hline
$\Lambda_b^0\to S_{-1/2}^{-} D^+ $ &
$\Xi_b^0\to S_{0}^- D^+ $ &
$\Xi_b^-\to S_{-1}^{--} D^+ $ &
$\Xi_b^0\to   F_{1/2}^-   D^+_s $ &
$\Xi_b^0\to   {F'}_0^-   D^+ $ \\

$\Sigma_{b}^{-}\to S_{-1}^{--} D^+_s $ &
$\Sigma_{b}^{0}\to   {F'}_0^-   D^+_s $ &
$\Xi_b^-\to   F_{-1/2}^{--}   D^+_s $ &
$\Sigma_{b}^{+}\to S_{1/2}^{0} D^+ $ &
$\Sigma_{b}^{0}\to S_{-1/2}^{-} D^+ $ \\
$\Sigma_{b}^{-}\to   F_{-3/2}^{--}   D^+ $ &
$\Sigma_{b}^{+}\to   F_1^{0}   D^+_s $ &
$\Sigma_{b}^{+}\to   {F'}_{1/2}^{0}   D^+ $ &
$\Sigma_{b}^{+}\to S_1^{0} D^+_s $ &
$\Sigma_{b}^{0}\to S_{0}^- D^+_s $ \\
$\Sigma_{b}^{0}\to   {F'}_{-1/2}^-   D^+ $&
$\Sigma_{b}^{-}\to   F_{-1}^{--}   D^+_s $ &
$\Sigma_{b}^{0}\to   F_{0}^-   D^+_s $ &
$\Sigma_{b}^{+}\to   T_{1/2}^{0}   D^+ $ &
$\Sigma_{b}^{0}\to   T_{-1/2}^-   D^+ $ \\
$\Sigma_{b}^{-}\to   T_{-3/2}^{--}   D^+ $ &
$\Sigma_{b}^{+}\to   T_{1}^{0}   D^+_s $ &
$\Sigma_{b}^{0}\to   T_0^-   D^+_s $ &
$\Xi_{b}^{\prime0}\to S_{0}^- D^+ $ &
$\Sigma_{b}^{-}\to   T_{-1}^{--}   D^+_s $ \\
$\Omega_{b}^{-}\to   F_{-1/2}^{--}   D^+ $ &
$\Xi_{b}^{\prime-}\to   F_{-1}^{--}   D^+ $ &
$\Xi_{b}^{\prime-}\to   F_{-1/2}^{--}   D^+_s $ &
$\Xi_{b}^{\prime0}\to   F_{1/2}^-  D^+_s $ &
$\Xi_{b}^{\prime0}\to   F_0^-   D^+ $ \\
$\Xi_{b}^{\prime0}\to   T_{1/2}^-   D^+_s $ &
$\Xi_{b}^{\prime-}\to   T_{-1/2}^{--}   D^+_s $ &
$\Omega_{b}^{-}\to   T_{0}^{--}   D^+_s $ &
$\Xi_{b}^{\prime-}\to S_{-1}^{--} D^+ $ &
$\Omega_{b}^{-}\to   T_{-1/2}^{--}   D^+ $ \\
$\Xi_{b}^{\prime0}\to   T_0^-   D^+ $ &
$\Xi_{b}^{\prime-}\to   T_{-1}^{--}   D^+ $ & &&\\\hline

%%%%%%%%%%%%%%%%%%%%%%%%%%%%%%%%%%%%%%%%%%%%%%%%%%%%%%%%%
%%%%%%%%%%%%%%%%%%%%%%%%%%%%%%%%%%%%%%%%%%%%%%%%%%%%%%%%%
$\Lambda_b^0\to   S_{-1/2}^{-}  \pi^+  $ &
$\Xi_b^0\to   S_{0}^-  \pi^+  $ &
$\Xi_b^-\to   S_1^{0}  \pi^-  $ &
$\Lambda_b^0\to   S_{0}^-  K^+  $&
$\Xi_b^-\to   S_{1/2}^{0}  K^-  $ \\
$\Lambda_b^0\to   F_1^{+}  K^-  $ &
$\Xi_b^-\to   S_{-1}^{--}  \pi^+  $ &
$\Lambda_b^0\to   {F'}_{-1/2}^-  \pi^+  $&
$\Xi_b^0\to   F_{1/2}^-  K^+  $ &
$\Lambda_b^0\to   F_0^-  K^+  $ \\
$\Xi_b^0\to   F_0^-  \pi^+  $ &
$\Xi_b^-\to   F_{-1}^{--}  \pi^+  $ &
$\Lambda_b^0\to   F_{3/2}^{+}  \pi^-  $ &
$\Xi_b^0\to   {F'}_0^-  \pi^+  $ &
$\Xi_b^-\to   {F'}_{1/2}^{0}  K^-  $ \\
$\Lambda_b^0\to   F_{-1/2}^-  \pi^+  $ &
$\Xi_b^0\to   F_{3/2}^{+}  K^-  $ &
$\Xi_b^-\to   F_1^{0}  \pi^-  $ &
$\Lambda_b^0\to   T_{3/2}^{+}  \pi^-  $ &
$\Xi_b^0\to   T_{3/2}^{+}  K^-  $ \\
$\Xi_b^-\to   T_{1/2}^{0}  K^-  $ &
$\Lambda_b^0\to   T_{-1/2}^-  \pi^+  $ &
$\Xi_b^0\to   T_0^-  \pi^+  $ &
$\Xi_b^-\to   T_{1}^{0}  \pi^-  $ &
$\Lambda_b^0\to   T_0^-  K^+  $ \\
$\Xi_b^0\to   T_{1/2}^-  K^+  $ &
$\Xi_b^-\to   F_{-1/2}^{--}  K^+  $ &
$\Sigma_{b}^{-}\to   S_{1/2}^{0}  \pi^-  $ &
$\Sigma_{b}^{+}\to   S_{1/2}^{0}  \pi^+  $ &
$\Sigma_{b}^{-}\to   S_0^{0}  K^-  $ \\
$\Sigma_{b}^{+}\to   S_1^{0}  K^+  $ &
$\Sigma_{b}^{0}\to   S_{-1/2}^{-}  \pi^+  $ &
$\Sigma_{b}^{-}\to   S_{-1}^{--}  K^+  $ &
$\Sigma_{b}^{0}\to   S_{0}^-  K^+  $ &
$\Sigma_{b}^{+}\to   F_{1/2}^{0}  \pi^+  $ \\
$\Sigma_{b}^{0}\to   T_{-1/2}^-  \pi^+  $ &
$\Sigma_{b}^{-}\to   F_0^{0}  K^-  $ &
$\Sigma_{b}^{+}\to   {F'}_{1/2}^{0}  \pi^+  $ &
$\Sigma_{b}^{0}\to   F_0^-  K^+  $ &
$\Sigma_{b}^{-}\to   F_{-1}^{--}  K^+  $ \\
$\Sigma_{b}^{+}\to   F_1^{0}  K^+  $ &
$\Sigma_{b}^{-}\to   {F'}_{1/2}^{0}  \pi^-  $ &
$\Sigma_{b}^{+}\to   T_2^{++}  K^-  $ &
$\Sigma_{b}^{0}\to   F_{-1/2}^-  \pi^+  $&
$\Sigma_{b}^{-}\to   F_{-3/2}^{--}  \pi^+  $ \\
$\Sigma_{b}^{+}\to   T_{1}^{0}  K^+  $ &
$\Sigma_{b}^{0}\to   T_{1}^{+}  K^-  $ &
$\Sigma_{b}^{-}\to   T_{0}^{0}  K^-  $ &
$\Sigma_{b}^{+}\to   T_{1/2}^{0}  \pi^+  $ &
$\Sigma_{b}^{0}\to   T_{3/2}^{+}  \pi^-  $ \\
$\Sigma_{b}^{-}\to   T_{1/2}^{0}  \pi^-  $ &
$\Sigma_{b}^{0}\to   F_{3/2}^{+}  \pi^-  $ &
$\Sigma_{b}^{-}\to   F_{1/2}^{0}  \pi^-  $ &
$\Sigma_{b}^{0}\to   T_0^-  K^+  $ &
$\Sigma_{b}^{-}\to   T_{-1}^{--}  K^+  $ \\
$\Sigma_{b}^{0}\to   {F'}_0^-  K^+  $ &
$\Sigma_{b}^{-}\to   T_{-3/2}^{--}  \pi^+  $ &
$\Sigma_{b}^{0}\to   {F'}_{-1/2}^-  \pi^+  $ &
$\Sigma_{b}^{0}\to   F_1^{+}  K^-  $ &$\Xi_{b}^{\prime-}\to   T_{-1/2}^{--}  K^+  $\\

%%%%%%%%%%%%%%%%%%%%%%%%%%%%%%%%%%%%%%%%%%%%%%%%%%%%%%%%%%
%%%%%%%%%%%%%%%%%%%%%%%%%%%%%%%%%%%%%%%%%%%%%%%%%%%%%%%%%%%%

$\Omega_{b}^{-}\to   S_1^{0}  K^-  $ &
$\Xi_{b}^{\prime0}\to   S_{0}^-  \pi^+  $ &
$\Xi_{b}^{\prime-}\to   S_1^{0}  \pi^-  $ &
$\Omega_{b}^{-}\to   F_{-1/2}^{--}  \pi^+  $ &
$\Xi_{b}^{\prime-}\to   S_{-1}^{--}  \pi^+  $ \\
$\Xi_{b}^{\prime-}\to   S_{1/2}^{0}  K^-  $ &
$\Omega_{b}^{-}\to   F_1^{0}  K^-  $ &
$\Omega_{b}^{-}\to   T_{1}^{0}  K^-  $ &
$\Xi_{b}^{\prime0}\to   F_{3/2}^{+}  K^-  $ &
$\Omega_{b}^{-}\to   T_{-1/2}^{--}  \pi^+  $ \\
$\Xi_{b}^{\prime0}\to   F_{1/2}^-  K^+  $ &
$\Xi_{b}^{\prime-}\to   F_{1/2}^{0}  K^-  $ &
$\Omega_{b}^{-}\to   T_{0}^{--}  K^+  $ &
$\Xi_{b}^{\prime0}\to   {F'}_0^-  \pi^+  $ &
$\Xi_{b}^{\prime0}\to   T_{3/2}^{+}  K^-  $ \\
$\Xi_{b}^{\prime-}\to   T_{1/2}^{0}  K^-  $ &
$\Xi_{b}^{\prime0}\to   F_0^-  \pi^+  $ &
$\Xi_{b}^{\prime-}\to   F_{-1}^{--}  \pi^+  $ &
$\Xi_{b}^{\prime0}\to   T_0^-  \pi^+  $ &
$\Xi_{b}^{\prime-}\to   T_{1}^{0}  \pi^-  $ \\
$\Xi_{b}^{\prime0}\to   T_{1/2}^-  K^+  $ &
$\Xi_{b}^{\prime-}\to   T_{-1}^{--}  \pi^+  $ &
$\Xi_{b}^{\prime-}\to   F_1^{0}  \pi^-  $ &
$\Xi_{b}^{\prime-}\to   F_{-1/2}^{--}  K^+  $ &
$\Xi_{b}^{\prime-}\to   {F'}_{1/2}^{0}  K^-  $ \\\hline

%%%%%%%%%%%%%%%%%%%%%%%%%%%%%%%%%

\end{tabular}
\end{table}
\section{Conclusion}
In conclusion, we discuss the production of singly charmed pentaquark $\bar c qqqq$, from the initial bottom baryon $bqq$, under the strategy of light quark SU(3) symmetry. The representation expressions of $\bar c qqqq$ are completed by the guidance of tensor reduction, formed by the projection of Eq.~\eqref{eq:tensor}, in particular collected in Appendix.~\ref{sec:app}. Following the constructed possible Hamiltonian of the production, we obtain cross sections of different production channels, as well as the relations between different channels. For definiteness, we suggest several golden channels for searching the singly charmed pentaquark in future experiments. The establishment the existence of these states means a remarkable progress in hadron physics.

\appendix
\section{tables and relations}
\label{sec:table}

We collect some results in the appendix. The tables of possible channels and amplitudes with the pentaquark multiplets $15$ and $15'$ are grouped into Tab~\ref{tab:P15_D}, Tab~\ref{tab:P15_M} and Tab~\ref{tab:P15p_DM}. Further more, for the relations between different channels, we deduce the cross section relations of $15$ state according to the amplitudes in Tab~\ref{tab:P15_D} and Tab~\ref{tab:P15_M}, shown as below.
\begin{small}
\begin{eqnarray*}
&&
\Gamma( \Lambda_b^0\to   F_0^{0}   D^0 ) =\Gamma( \Lambda_b^0\to   F_{-1}^-   D^+ ) =
\Gamma( \Lambda_b^0\to   {F'}_{-1/2}^-   D^+_s )=\Gamma( \Xi_b^-\to   F_{-1/2}^-   D^0 )=\Gamma( \Xi_b^-\to   F_{-3/2}^{--}   D^+ )\\&&=\Gamma( \Xi_b^-\to   F_{-1}^{--}   D^+_s )=
\Gamma( \Xi_b^0\to   F_{1/2}^{0}   D^0 ) =\Gamma( \Xi_b^0\to   {F'}_0^-   D^+_s ),\\
&&\Gamma( \Lambda_b^0\to   {F'}_{1/2}^{0}   D^0 ) =\Gamma( \Lambda_b^0\to   {F'}_{-1/2}^-   D^+ ) =
\Gamma( \Xi_b^-\to   F_0^-   D^0 ) =\Gamma( \Xi_b^-\to   F_{-1}^{--}   D^+	)=\Gamma( \Xi_b^-\to   F_{-1/2}^{--}   D^+_s	)\\&&=\Gamma( \Xi_b^0\to   F_1^{0}   D^0 )=
\Gamma( \Xi_b^0\to   {F'}_0^-   D^+ ) =\Gamma( \Xi_b^0\to   F_{1/2}^-   D^+_s),\\
&&
\Gamma( \Xi_b^-\to   F_{1/2}^-  K^0 ) =\Gamma( \Xi_b^-\to   {F'}_{-1/2}^-  \overline K^0 ) =
6 \Gamma( \Xi_b^-\to   {F'}_0^-  \eta_q ) =2 \Gamma( \Xi_b^-\to   {F'}_0^-  \pi^0 ),\\
&&
\Gamma( \Lambda_b^0\to   F_{-1}^-  \pi^+) =\Gamma( \Lambda_b^0\to   {F'}_{-1/2}^-  K^+),\quad \Gamma( \Lambda_b^0\to   F_{1/2}^{0}  K^0) =\Gamma( \Lambda_b^0\to   F_{-1/2}^-  K^+),\\
&&
\Gamma( \Xi_b^0\to   F_1^{+}  K^-) =\Gamma( \Xi_b^0\to   F_0^-  K^+),\quad \Gamma( \Xi_b^-\to   F_{-3/2}^{--}  \pi^+ ) =\Gamma( \Xi_b^-\to   F_{-1}^{--}  K^+ ),\\
&&
\Gamma( \Xi_b^0\to   F_0^{0}  \overline K^0) =\Gamma( \Xi_b^0\to   F_{-1/2}^-  \pi^+),\quad \Gamma( \Xi_b^0\to   F_{3/2}^{+}  \pi^-) =\Gamma( \Xi_b^0\to   F_1^{0}  K^0),\\
&&
\Gamma( \Xi_b^-\to   F_0^{0}  K^-) =\Gamma( \Xi_b^-\to   F_{1/2}^{0}  \pi^-),\quad \Gamma( \Xi_b^-\to   F_{-1}^{--}  \pi^+ ) =\Gamma( \Xi_b^-\to   F_{-1/2}^{--}  K^+ ), \\
&&
\Gamma( \Lambda_b^0\to   {F'}_{-1/2}^-  \pi^+ ) =\Gamma( \Lambda_b^0\to   {F'}_{1/2}^{0}  \pi^0 ),\quad \Gamma( \Lambda_b^0\to   F_1^{+}  K^- ) =\Gamma( \Lambda_b^0\to   F_0^{0}  \overline K^0 ),\\
&&
\Gamma( \Lambda_b^0\to   F_1^{0}  K^0 ) =\Gamma( \Lambda_b^0\to   F_0^-  K^+ ),\quad \Gamma( \Xi_b^0\to   F_{1/2}^-  K^+ ) =\Gamma( \Xi_b^0\to   {F'}_0^-  \pi^+ ),\\
&&
\Gamma( \Xi_b^0\to   F_0^-  \pi^+ ) =\Gamma( \Xi_b^0\to   {F'}_{1/2}^{0}  \overline K^0 ),\quad \Gamma( \Xi_b^-\to   {F'}_{1/2}^{0}  K^- ) =\Gamma( \Xi_b^-\to   F_1^{0}  \pi^- ),\\
&&
\Gamma( \Lambda_b^0\to   F_1^{+}  \pi^- ) =\Gamma( \Lambda_b^0\to   {F'}_{1/2}^{0}  K^0 ) =\Gamma( \Xi_b^-\to   F_0^-  K^0 ),\\
&&
\Gamma( \Xi_b^-\to   F_{-1}^-  \overline K^0) =\Gamma( \Xi_b^-\to   {F'}_0^-  K^0) =\Gamma( \Xi_b^-\to   {F'}_{-1/2}^-  \eta_q),\\
&&
\Gamma( \Lambda_b^0\to   F_{3/2}^{+}  \pi^-) =\Gamma( \Lambda_b^0\to   F_{-1/2}^-  \pi^+) =6 \Gamma( \Lambda_b^0\to   F_{1/2}^{0}  \eta_q ),\\
&&
\Gamma( \Xi_b^-\to   F_{-1/2}^-  \overline K^0 ) =\Gamma( \Xi_b^0\to   F_{1/2}^{0}  \overline K^0 ) =
\Gamma( \Xi_b^0\to   F_{3/2}^{+}  K^- ),\\
&&
2 \Gamma( \Sigma_{b}^{0}\to   F_{-1}^-   D^+) =2 \Gamma( \Xi_{b}^{\prime0}\to   F_{1/2}^{0}   D^0) =
2 \Gamma( \Xi_{b}^{\prime0}\to   {F'}_{-1/2}^-   D^+) =2 \Gamma( \Xi_{b}^{\prime0}\to   F_{-1/2}^-   D^+) =2 \Gamma( \Xi_{b}^{\prime0}\to   {F'}_0^-   D^+_s) \\
&&=2 \Gamma( \Xi_{b}^{\prime-}\to   F_{-3/2}^{--}   D^+) =\Gamma( \Omega_{b}^{-}\to   {F'}_0^-   D^0)=\Gamma( \Sigma_{b}^{+}\to   F_0^{0}   D^+)=\Gamma( \Sigma_{b}^{+}\to   F_{1/2}^{0}   D^+_s)
=\Gamma( \Omega_{b}^{-}\to   F_{-1}^{--}   D^+),\\
&&
\Gamma( \Sigma_{b}^{0}\to   {F'}_{-1/2}^-   D^+) =\Gamma( \Sigma_{b}^{0}\to   F_0^-   D^+_s) =
\Gamma( \Xi_{b}^{\prime-}\to   F_{-1/2}^{--}   D^+_s) =\Gamma( \Sigma_{b}^{0}\to   {F'}_{1/2}^{0}   D^0)=\Gamma( \Sigma_{b}^{0}\to   {F'}_0^-   D^+_s) \\
&&=\Gamma( \Xi_{b}^{\prime0}\to   F_{1/2}^-   D^+_s) =
2 \Gamma( \Sigma_{b}^{+}\to   {F'}_{1/2}^{0}   D^+) =2 \Gamma( \Sigma_{b}^{-}\to   F_{-1}^{--}   D^+_s)=2 \Gamma( \Sigma_{b}^{+}\to   F_1^{0}   D^+_s)=2 \Gamma( \Sigma_{b}^{-}\to   {F'}_{-1/2}^-   D^0),\\
&&
2 \Gamma( \Sigma_{b}^{-}\to   F_{-3/2}^{--}   D^+) =\Gamma( \Sigma_{b}^{0}\to   F_{-1/2}^-   D^+) =
2 \Gamma( \Sigma_{b}^{+}\to   F_{3/2}^{+}   D^0) =\Gamma( \Sigma_{b}^{0}\to   F_{1/2}^{0}   D^0),\\
&&
\Gamma( \Sigma_{b}^{0}\to   F_0^{0}   D^0) =\Gamma( \Xi_{b}^{\prime-}\to   {F'}_{-1/2}^-   D^0) =
\Gamma( \Xi_{b}^{\prime-}\to   F_{-1}^{--}   D^+_s)=\Gamma( \Sigma_{b}^{0}\to   F_{-1/2}^-   D^+_s),\\
&&
\Gamma( \Xi_{b}^{\prime0}\to   F_0^-   D^+) =\Gamma( \Xi_{b}^{\prime-}\to   F_{-1}^{--}   D^+) =
\Gamma( \Xi_{b}^{\prime0}\to   F_1^{0}   D^0) =\Gamma( \Xi_{b}^{\prime-}\to   {F'}_0^-   D^0),\\
&&\Gamma( \Xi_{b}^{\prime0}\to   {F'}_{1/2}^{0}   D^0 ) =2 \Gamma( \Sigma_{b}^{+}\to   F_1^{+}   D^0 ) =
\Gamma( \Xi_{b}^{\prime0}\to   F_0^-   D^+_s ) =2 \Gamma( \Omega_{b}^{-}\to   F_{-1/2}^{--}   D^+_s ),\\
&&
\Gamma( \Sigma_{b}^{-}\to   F_{-1}^-   D^0) =\Gamma( \Sigma_{b}^{-}\to   F_{-3/2}^{--}   D^+_s),\quad \Gamma( \Omega_{b}^{-}\to   F_{-1/2}^{--}   D^+) =\Gamma( \Omega_{b}^{-}\to   F_{1/2}^-   D^0),\\
&&
\Gamma( \Omega_{b}^{-}\to   F_{1/2}^{0}  K^- ) =\Gamma( \Omega_{b}^{-}\to   F_{-1/2}^-  \overline K^0 ),\quad \Gamma( \Sigma_{b}^{-}\to   F_{-1/2}^-  K^0 ) =\Gamma( \Sigma_{b}^{+}\to   F_{1/2}^{0}  K^+ ),\\
&&
\Gamma( \Sigma_{b}^{-}\to   F_0^{0}  K^- ) =2 \Gamma( \Sigma_{b}^{0}\to   F_1^{+}  K^- ),\quad \Gamma( \Omega_{b}^{-}\to   {F'}_0^-  \overline K^0 ) =\Gamma( \Sigma_{b}^{+}\to   F_{1/2}^{0}  \pi^+ ),\\
&&
\Gamma( \Sigma_{b}^{+}\to   {F'}_{1/2}^{0}  \pi^+ ) =\Gamma( \Omega_{b}^{-}\to   F_0^-  \overline K^0 ),\quad \Gamma( \Omega_{b}^{-}\to   F_{-1/2}^{--}  \pi^+ ) =2 \Gamma( \Omega_{b}^{-}\to   F_{1/2}^-  \pi^0 ),\\
&&
\Gamma( \Sigma_{b}^{-}\to   F_{-1}^{--}  K^+ ) =2 \Gamma( \Xi_{b}^{\prime-}\to   F_{-1/2}^{--}  K^+ ),\quad \Gamma( \Sigma_{b}^{-}\to   {F'}_{-1/2}^-  \pi^0 ) =\Gamma( \Xi_{b}^{\prime-}\to   F_{-1/2}^-  \overline K^0 ),\\
&&
\Gamma( \Sigma_{b}^{0}\to   F_0^-  K^+ ) =\Gamma( \Xi_{b}^{\prime-}\to   {F'}_{1/2}^{0}  K^- ),\quad \Gamma( \Sigma_{b}^{0}\to   {F'}_{1/2}^{0}  \eta_q ) =2 \Gamma( \Sigma_{b}^{-}\to   {F'}_{-1/2}^-  \eta_q ),\\
&&
\Gamma( \Omega_{b}^{-}\to   F_{-1}^{--}  \pi^+ ) =2 \Gamma( \Xi_{b}^{\prime-}\to   F_{-3/2}^{--}  \pi^+ ) =
2 \Gamma( \Omega_{b}^{-}\to   F_0^-  \pi^0 ),\\
&&
\Gamma( \Omega_{b}^{-}\to   F_1^{0}  \pi^- ) =2 \Gamma( \Xi_{b}^{\prime0}\to   F_{3/2}^{+}  \pi^- ) =
2 \Gamma( \Omega_{b}^{-}\to   {F'}_0^-  \pi^0 ),\\
&&
\Gamma( \Sigma_{b}^{-}\to   F_0^-  K^0 ) =\Gamma( \Sigma_{b}^{-}\to   {F'}_{1/2}^{0}  \pi^- ).
\end{eqnarray*}
\end{small}
The cross section relations of $15'$ state are given as follows.
\begin{small}
\begin{eqnarray*}
&&
\Gamma( \Lambda_b^0\to   T_{1}^{+}  \pi^- ) =\Gamma( \Lambda_b^0\to   T_{1/2}^{0}  K^0 ) =
\Gamma( \Xi_b^0\to   T_{1}^{+}  K^- ) =\Gamma( \Xi_b^0\to   T_{0}^{0}  \overline K^0 ) =\Gamma( \Xi_b^-\to   T_{0}^{0}  K^- )  \\
&&=\Gamma( \Xi_b^-\to   T_0^-  K^0 ) =\Gamma( \Xi_b^0\to   T_{-1/2}^-  \pi^+ ) =\Gamma( \Xi_b^0\to   T_0^-  K^+ ) =\Gamma( \Xi_b^-\to   T_{1/2}^{0}  \pi^- ) =
\Gamma( \Lambda_b^0\to   T_{-1}^-  \pi^+ )\\
&& =\Gamma( \Lambda_b^0\to   T_{-1/2}^-  K^+ ) =2 \Gamma( \Xi_b^0\to   T_{1/2}^{0}  \pi^0 ) =2 \Gamma( \Xi_b^-\to   T_{-1/2}^-  \pi^0 ) =
\frac{1}{2} \Gamma( \Lambda_b^0\to   T_{0}^{0}  \pi^0 )=\frac{2}{3} \Gamma( \Xi_b^0\to   T_{1/2}^{0}  \eta_q ) \\
&&=\frac{2}{3} \Gamma( \Xi_b^-\to   T_{-1/2}^-  \eta_q )=\Gamma( \Xi_b^-\to   T_{-1}^-  \overline K^0 ),\\
&&
\Gamma( \Lambda_b^0\to   T_{3/2}^{+}  \pi^- ) =\Gamma( \Xi_b^-\to   T_{1/2}^{0}  K^- ) =
\Gamma( \Xi_b^-\to   T_{-1/2}^-  \overline K^0 )=\Gamma( \Lambda_b^0\to   T_{1}^{0}  K^0 ) =\Gamma( \Xi_b^0\to   T_{3/2}^{+}  K^- ) \\
&&=\Gamma( \Xi_b^0\to   T_{1/2}^{0}  \overline K^0 ) =
\Gamma( \Xi_b^0\to   T_0^-  \pi^+ ) =\Gamma( \Xi_b^0\to   T_{1/2}^-  K^+ ) =\Gamma( \Xi_b^-\to   T_{1}^{0}  \pi^- ) =\Gamma( \Xi_b^-\to   T_{1/2}^-  K^0 ) \\
&&=
\Gamma( \Lambda_b^0\to   T_{-1/2}^-  \pi^+ ) =\Gamma( \Lambda_b^0\to   T_0^-  K^+ ) =2 \Gamma( \Xi_b^0\to   T_{1}^{0}  \pi^0 ) =2 \Gamma( \Xi_b^-\to   T_0^-  \pi^0 ) =
\frac{1}{2} \Gamma( \Lambda_b^0\to   T_{1/2}^{0}  \pi^0 )\\
&& =\frac{2}{3} \Gamma( \Xi_b^-\to   T_0^-  \eta_q ) =\frac{2}{3} \Gamma( \Xi_b^0\to   T_{1}^{0}  \eta_q ),\\
&&
\Gamma( \Sigma_{b}^{0}\to   T_{0}^{0}   D^0) =\Gamma( \Sigma_{b}^{0}\to   T_{-1}^-   D^+) =
\Gamma( \Sigma_{b}^{0}\to   T_{-1/2}^-   D^+_s) =\Gamma( \Xi_{b}^{\prime0}\to   T_{1/2}^{0}   D^0)=\Gamma( \Xi_{b}^{\prime0}\to   T_{-1/2}^-   D^+) \\
&&=\Gamma( \Xi_{b}^{\prime0}\to   T_0^-   D^+_s)=
\Gamma( \Xi_{b}^{\prime-}\to   T_{-1/2}^-   D^0) =\Gamma( \Xi_{b}^{\prime-}\to   T_{-3/2}^{--}   D^+) =\Gamma( \Xi_{b}^{\prime-}\to   T_{-1}^{--}   D^+_s) =2 \Gamma( \Sigma_{b}^{+}\to   T_{1}^{+}   D^0) \\
&&=2 \Gamma( \Sigma_{b}^{+}\to   T_{0}^{0}   D^+) =2 \Gamma( \Sigma_{b}^{+}\to   T_{1/2}^{0}   D^+_s) =2 \Gamma( \Sigma_{b}^{-}\to   T_{-1}^-   D^0) =2 \Gamma( \Sigma_{b}^{-}\to   T_{-2}^{--}   D^+) =
2 \Gamma( \Sigma_{b}^{-}\to   T_{-3/2}^{--}   D^+_s) \\
&&=2 \Gamma( \Omega_{b}^{-}\to   T_0^-   D^0)=2 \Gamma( \Omega_{b}^{-}\to   T_{-1}^{--}   D^+) =2 \Gamma( \Omega_{b}^{-}\to   T_{-1/2}^{--}   D^+_s),\\
&&
\Gamma( \Xi_{b}^{\prime0}\to   T_{1}^{0}   D^0) =\Gamma( \Sigma_{b}^{0}\to   T_0^-   D^+_s) =
\Gamma( \Xi_{b}^{\prime-}\to   T_0^-   D^0)=2 \Gamma( \Omega_{b}^{-}\to   T_{-1/2}^{--}   D^+) =2 \Gamma( \Omega_{b}^{-}\to   T_{0}^{--}   D^+_s) \\
&&=\Gamma( \Xi_{b}^{\prime-}\to   T_{-1/2}^{--}   D^+_s) =\Gamma( \Xi_{b}^{\prime-}\to   T_{-1}^{--}   D^+)=\Gamma( \Sigma_{b}^{0}\to   T_{-1/2}^-   D^+)=\Gamma( \Sigma_{b}^{0}\to   T_{1/2}^{0}   D^0) =2 \Gamma( \Sigma_{b}^{+}\to   T_{3/2}^{+}   D^0)\\
&& =
2 \Gamma( \Sigma_{b}^{+}\to   T_{1/2}^{0}   D^+) =2 \Gamma( \Sigma_{b}^{+}\to   T_{1}^{0}   D^+_s) =2 \Gamma( \Sigma_{b}^{-}\to   T_{-1/2}^-   D^0) =2 \Gamma( \Sigma_{b}^{-}\to   T_{-3/2}^{--}   D^+)=\Gamma( \Xi_{b}^{\prime0}\to   T_0^-   D^+)  \\
&&=2 \Gamma( \Sigma_{b}^{-}\to   T_{-1}^{--}   D^+_s) =2 \Gamma( \Omega_{b}^{-}\to   T_{1/2}^-   D^0) =\Gamma( \Xi_{b}^{\prime0}\to   T_{1/2}^-   D^+_s),\\
&&
\Gamma( \Xi_{b}^{\prime-}\to   T_{-1}^{--}  \pi^+ ) =\Gamma( \Xi_{b}^{\prime-}\to   T_{-1/2}^{--}  K^+ ) =
2 \Gamma( \Sigma_{b}^{0}\to   T_{1/2}^{0}  \pi^0 ) =2 \Gamma( \Sigma_{b}^{-}\to   T_{-3/2}^{--}  \pi^+ )=4 \Gamma( \Omega_{b}^{-}\to   T_{1/2}^-  \pi^0 ) \\
&&=2 \Gamma( \Sigma_{b}^{-}\to   T_{-1}^{--}  K^+ ) =2 \Gamma( \Omega_{b}^{-}\to   T_{-1/2}^{--}  \pi^+ ) =
2 \Gamma( \Omega_{b}^{-}\to   T_{0}^{--}  K^+ ) ,\\
&&
\Gamma( \Sigma_{b}^{0}\to   T_{1}^{+}  K^- ) =\Gamma( \Sigma_{b}^{0}\to   T_{0}^{0}  \overline K^0 ) =
2 \Gamma( \Sigma_{b}^{+}\to   T_2^{++}  K^- ) =2 \Gamma( \Sigma_{b}^{-}\to   T_{-1}^-  \overline K^0 )=2 \Gamma( \Sigma_{b}^{+}\to   T_{1}^{+}  \overline K^0 )  \\
&&=2 \Gamma( \Sigma_{b}^{-}\to   T_{0}^{0}  K^- ),\\
&&
\Gamma( \Sigma_{b}^{-}\to   T_{1/2}^{0}  \pi^- ) =\Gamma( \Sigma_{b}^{-}\to   T_0^-  K^0 ) =
2 \Gamma( \Sigma_{b}^{0}\to   T_{3/2}^{+}  \pi^- ) =2 \Gamma( \Sigma_{b}^{0}\to   T_{1}^{0}  K^0 )=2 \Gamma( \Xi_{b}^{\prime-}\to   T_{1}^{0}  \pi^- ) \\
&& =2 \Gamma( \Xi_{b}^{\prime-}\to   T_{1/2}^-  K^0 ),\\
&&
\Gamma( \Sigma_{b}^{0}\to   T_{-1/2}^-  \pi^+ ) =\Gamma( \Sigma_{b}^{0}\to   T_0^-  K^+ ) =
\Gamma( \Xi_{b}^{\prime0}\to   T_0^-  \pi^+ ) =\Gamma( \Xi_{b}^{\prime0}\to   T_{1/2}^-  K^+ ) =2 \Gamma( \Xi_{b}^{\prime0}\to   T_{1}^{0}  \pi^0 ),\\
&&
\Gamma( \Omega_{b}^{-}\to   T_{1/2}^{0}  K^- ) =\Gamma( \Omega_{b}^{-}\to   T_{-1/2}^-  \overline K^0 ) =
2 \Gamma( \Xi_{b}^{\prime0}\to   T_{1}^{+}  K^- ) =2 \Gamma( \Xi_{b}^{\prime0}\to   T_{0}^{0}  \overline K^0 )=2 \Gamma( \Xi_{b}^{\prime-}\to   T_{0}^{0}  K^- )\\
&& =2 \Gamma( \Xi_{b}^{\prime-}\to   T_{-1}^-  \overline K^0 ),\\
&&
\Gamma( \Xi_{b}^{\prime0}\to   T_{3/2}^{+}  \pi^- ) =\Gamma( \Xi_{b}^{\prime0}\to   T_{1}^{0}  K^0 ) =
2 \Gamma( \Sigma_{b}^{+}\to   T_2^{++}  \pi^- ) =2 \Gamma( \Sigma_{b}^{+}\to   T_{3/2}^{+}  K^0 )=2 \Gamma( \Omega_{b}^{-}\to   T_{1}^{0}  \pi^- ) \\
&& =2 \Gamma( \Omega_{b}^{-}\to   T_{1/2}^-  K^0 ),\\
&&
\Gamma( \Xi_{b}^{\prime-}\to   T_{-3/2}^{--}  \pi^+ ) =\Gamma( \Xi_{b}^{\prime-}\to   T_{-1}^{--}  K^+ ) =
2 \Gamma( \Sigma_{b}^{-}\to   T_{-2}^{--}  \pi^+ ) =2 \Gamma( \Sigma_{b}^{-}\to   T_{-3/2}^{--}  K^+ ) \\
&&=2 \Gamma( \Omega_{b}^{-}\to   T_{-1}^{--}  \pi^+ ) =2 \Gamma( \Omega_{b}^{-}\to   T_{-1/2}^{--}  K^+ ),\\
&&
\Gamma( \Xi_{b}^{\prime0}\to   T_{3/2}^{+}  K^- ) =\Gamma( \Xi_{b}^{\prime0}\to   T_{1/2}^{0}  \overline K^0 ) =
\Gamma( \Xi_{b}^{\prime-}\to   T_{1/2}^{0}  K^- ) =\Gamma( \Xi_{b}^{\prime-}\to   T_{-1/2}^-  \overline K^0 ),\\
&&
\Gamma( \Sigma_{b}^{0}\to   T_{1}^{+}  \pi^- ) =\Gamma( \Sigma_{b}^{0}\to   T_{1/2}^{0}  K^0 ) =
\Gamma( \Xi_{b}^{\prime-}\to   T_{1/2}^{0}  \pi^- ) =\Gamma( \Xi_{b}^{\prime-}\to   T_0^-  K^0 ),\\
&&
\Gamma( \Sigma_{b}^{0}\to   T_{-1}^-  \pi^+ ) =\Gamma( \Xi_{b}^{\prime0}\to   T_{-1/2}^-  \pi^+ ) =
\Gamma( \Sigma_{b}^{0}\to   T_{-1/2}^-  K^+ ) =\Gamma( \Xi_{b}^{\prime0}\to   T_0^-  K^+ ),\\
&&\Gamma( \Xi_{b}^{\prime0}\to   T_{1}^{0}  \eta_q ) =
\Gamma( \Xi_{b}^{\prime-}\to   T_0^-  \eta_q ),\quad \Gamma( \Omega_{b}^{-}\to   T_{1}^{0}  K^- ) =
\Gamma( \Omega_{b}^{-}\to   T_0^-  \overline K^0 ),\\
&&\Gamma( \Sigma_{b}^{0}\to   T_{0}^{0}  \pi^0 ) =2 \Gamma( \Omega_{b}^{-}\to   T_0^-  \pi^0 ),\quad \Gamma( \Sigma_{b}^{-}\to   T_{0}^{0}  \pi^- ) =
\Gamma( \Sigma_{b}^{-}\to   T_{-1/2}^-  K^0 ),\\
&&
\Gamma( \Sigma_{b}^{+}\to   T_{1}^{+}  \eta_q ) =\Gamma( \Sigma_{b}^{-}\to   T_{-1}^-  \eta_q ),\quad \Gamma( \Sigma_{b}^{+}\to   T_{0}^{0}  \pi^+ ) =
\Gamma( \Sigma_{b}^{+}\to   T_{1/2}^{0}  K^+ ),\\
&&
\Gamma( \Sigma_{b}^{+}\to   T_{1/2}^{0}  \pi^+ ) =\Gamma( \Sigma_{b}^{+}\to   T_{1}^{0}  K^+ ) =
2 \Gamma( \Sigma_{b}^{+}\to   T_{3/2}^{+}  \pi^0 ),\\
&&
\Gamma( \Sigma_{b}^{0}\to   T_{1/2}^{0}  \eta_q ) =2 \Gamma( \Sigma_{b}^{+}\to   T_{3/2}^{+}  \eta_q ) =
2 \Gamma( \Sigma_{b}^{-}\to   T_{-1/2}^-  \eta_q ),\\
&&\Gamma( \Xi_{b}^{\prime0}\to   T_{1/2}^{0}  \eta_q ) =\Gamma( \Xi_{b}^{\prime-}\to   T_{-1/2}^-  \eta_q ).
\end{eqnarray*}
\end{small}

\begin{table}
\tiny
\caption{The productions of pentaquark $P_{15}$ from singly bottom baryons $T_{b\bar3}/T_{b6}$, with one meson (charmed meson or light meson) in the final state.}\label{tab:P15_D}
\begin{tabular}{|lr|lr|lr|}\hline\hline
channel & amplitude &channel &amplitude&channel &amplitude \\\hline
$\Lambda_b^0\to   F_0^{0}   D^0 $ & $ b_1 V_{cd}^*$&
$\Xi_b^-\to   F_{-3/2}^{--}   D^+ $ & $ b_1 V_{cd}^*$&
$\Xi_b^0\to   {F'}_0^-   D^+_s $ & $ -b_1 V_{cd}^*$\\

$\Lambda_b^0\to   F_{-1}^-   D^+ $ & $ b_1 V_{cd}^*$&
$\Xi_b^-\to   F_{-1}^{--}   D^+/D^+_s $ & $ b_1 V_{cs}^*/b_1 V_{cd}^*$&
$\Xi_b^0\to   F_1^{0}   D^0 $ & $ -b_1 V_{cs}^*$\\

$\Lambda_b^0\to   {F'}_{-1/2}^-   D^+/D^+_s $ & $ b_1 V_{cs}^*/b_1 V_{cd}^*$&
$\Xi_b^-\to   F_{-1/2}^{--}   D^+_s $ & $ b_1 V_{cs}^*$&
$\Xi_b^0\to   F_{1/2}^{0}   D^0 $ & $ -b_1 V_{cd}^*$\\

$\Lambda_b^0\to   {F'}_{1/2}^{0}   D^0 $ & $ b_1 V_{cs}^*$&
$\Xi_b^-\to   F_0^-   D^0 $ & $ b_1 V_{cs}^*$&
$\Xi_b^0\to   {F'}_0^-   D^+ $ & $ -b_1 V_{cs}^*$\\

$\Lambda_b^0\to   F_{-1}^-  \pi^+  $ & $ (e_4-e_2) V_{cd}^*$&
$\Xi_b^-\to   F_{-1/2}^-   D^0 $ & $ b_1 V_{cd}^*$&
$\Xi_b^0\to   F_{1/2}^-   D^+_s $ & $ -b_1 V_{cs}^*$\\

$\Lambda_b^0\to   {F'}_{-1/2}^-  K^+  $ & $ (e_4-e_2) V_{cd}^*$&
$\Xi_b^-\to   F_{-1}^-  \overline K^0  $ & $ (e_1+e_3) V_{cd}^*$&
$\Xi_b^0\to   F_1^{+}  K^-  $ & $ (e_1-e_3) V_{cd}^*$\\

$\Lambda_b^0\to   F_{1/2}^{0}  K^0  $ & $ 2 e_5 V_{cd}^*$&
$\Xi_b^-\to   {F'}_{-1/2}^-  \eta_q  $ & $ -\sqrt{\frac{2}{3}} (e_1+e_3) V_{cd}^*$&
$\Xi_b^0\to   F_0^-  K^+  $ & $ (e_1-e_3) V_{cd}^*$\\

$\Lambda_b^0\to   F_{-1/2}^-  K^+  $ & $ 2 e_5 V_{cd}^*$&
$\Xi_b^-\to   {F'}_0^-  K^0  $ & $ (e_1+e_3) V_{cd}^*$&
$\Xi_b^0\to   F_0^{0}  \overline K^0  $ & $ (e_1-e_3-2 e_5) V_{cd}^*$\\

$\Lambda_b^0\to   F_1^{+}  \pi^-  $ & $ (e_2+e_4) V_{cd}^*$&
$\Xi_b^-\to   F_0^-  K^0  $ & $ (e_2+e_4) V_{cd}^*$&
$\Xi_b^0\to   F_{-1/2}^-  \pi^+  $ & $ (e_1-e_3-2 e_5) V_{cd}^*$\\

$\Lambda_b^0\to   {F'}_{1/2}^{0}  K^0  $ & $ (e_2+e_4) V_{cd}^*$&
$\Xi_b^-\to   F_{-3/2}^{--}  \pi^+  $ & $ (e_1-e_2+e_3+e_4) V_{cd}^*$&
$\Xi_b^0\to   F_{3/2}^{+}  \pi^-  $ & $ (e_1-e_2-e_3-e_4) V_{cd}^*$\\

$\Lambda_b^0\to   F_0^{0}  \overline K^0  $ & $ (e_2+e_3+e_4-e_1) V_{cs}^*$&
$\Xi_b^-\to   F_{-1}^{--}  K^+  $ & $ (e_1-e_2+e_3+e_4) V_{cd}^*$&
$\Xi_b^0\to   F_1^{0}  K^0  $ & $ (e_1-e_2-e_3-e_4) V_{cd}^*$\\

$\Lambda_b^0\to   F_1^{+}  K^-  $ & $ (e_2+e_3+e_4-e_1) V_{cs}^*$&
$\Xi_b^-\to   F_0^{0}  K^-  $ & $ (e_1+e_3+2 e_5) V_{cd}^*$&
$\Xi_b^0\to   {F'}_0^-  K^+  $ & $ (e_2-e_4) V_{cd}^*$\\

$\Lambda_b^0\to   {F'}_{1/2}^{0}  \pi^0  $ & $ \frac{(e_4-e_2) V_{cs}^*}{\sqrt{2}}$&
$\Xi_b^-\to   F_{1/2}^{0}  \pi^-  $ & $ (e_1+e_3+2 e_5) V_{cd}^*$&
$\Xi_b^0\to   {F'}_0^-  \pi^+  $ & $ (e_2-e_4) V_{cs}^*$\\

$\Lambda_b^0\to   {F'}_{-1/2}^-  \pi^+  $ & $ (e_4-e_2) V_{cs}^*$&
$\Xi_b^-\to   F_{-1/2}^-  \pi^0  $ & $ \frac{(e_1-2 e_2+e_3+2 e_5) V_{cd}^*}{\sqrt{2}}$&
$\Xi_b^0\to   F_{1/2}^-  K^+  $ & $ (e_2-e_4) V_{cs}^*$\\

$\Lambda_b^0\to   F_1^{0}  K^0  $ & $ (-e_1+e_3+2 e_5) V_{cs}^*$&
$\Xi_b^-\to   F_{-1/2}^-  \eta_q  $ & $ \frac{(e_1+e_3+2 e_4+2 e_5) V_{cd}^*}{\sqrt{6}}$&
$\Xi_b^0\to   {F'}_{1/2}^{0}  \overline K^0  $ & $ -2 e_5 V_{cs}^*$\\

$\Lambda_b^0\to   F_0^-  K^+  $ & $ (e_3+2 e_5-e_1) V_{cs}^*$&
$\Xi_b^-\to   {F'}_0^-  \pi^0  $ & $ -\frac{(e_1+e_3) V_{cs}^*}{\sqrt{2}}$&
$\Xi_b^0\to   F_0^-  \pi^+  $ & $ -2 e_5 V_{cs}^*$\\

$\Lambda_b^0\to   F_0^{0}  \eta_q  $ & $ \sqrt{\frac{2}{3}} (e_4-2 e_5) V_{cd}^*$&
$\Xi_b^-\to   {F'}_0^-  \eta_q  $ & $ \frac{(e_1+e_3) V_{cs}^*}{\sqrt{6}}$&
$\Xi_b^0\to   F_{3/2}^{+}  K^-  $ & $ -(e_2+e_4) V_{cs}^*$\\

$\Lambda_b^0\to   F_0^{0}  \pi^0  $ & $ -\sqrt{2} e_2 V_{cd}^*$&
$\Xi_b^-\to   {F'}_{-1/2}^-  \overline K^0  $ & $ (e_1+e_3) V_{cs}^*$&
$\Xi_b^0\to   {F'}_{1/2}^{0}  \eta_q  $ & $ \sqrt{\frac{2}{3}} (e_3-e_1) V_{cd}^*$\\

$\Lambda_b^0\to   F_{3/2}^{+}  \pi^-  $ & $ (e_3-e_1) V_{cs}^*$&
$\Xi_b^-\to   F_{1/2}^-  K^0  $ & $ (e_1+e_3) V_{cs}^*$&
$\Xi_b^0\to   F_{1/2}^{0}  \overline K^0  $ & $ -(e_2+e_4) V_{cs}^*$\\

$\Lambda_b^0\to   F_{1/2}^{0}  \eta_q  $ & $ \frac{(e_3-e_1) V_{cs}^*}{\sqrt{6}}$&
$\Xi_b^-\to   {F'}_{1/2}^{0}  K^-  $ & $ (e_1+e_3+2 e_5) V_{cs}^*$&
$\Xi_b^0\to   F_1^{0}  \pi^0  $ & $ \frac{(e_2-e_4+2 e_5) V_{cs}^*}{\sqrt{2}}$\\

$\Lambda_b^0\to   F_{-1/2}^-  \pi^+  $ & $ (e_3-e_1) V_{cs}^*$&
$\Xi_b^-\to   F_1^{0}  \pi^-  $ & $ (e_1+e_3+2 e_5) V_{cs}^*$&
$\Xi_b^0\to   F_1^{0}  \eta_q  $ & $ \frac{(3 e_2+e_4-2 e_5) V_{cs}^*}{\sqrt{6}}$\\

$\Lambda_b^0\to   F_{1/2}^{0}  \pi^0  $ & $ \frac{(e_1-e_3) V_{cs}^*}{\sqrt{2}}$&
$\Xi_b^-\to   F_{-1}^{--}  \pi^+  $ & $ (e_1-e_2+e_3+e_4) V_{cs}^*$&
$\Xi_b^0\to   F_{1/2}^{0}  \pi^0  $ & $ \frac{(2 e_2-e_1+e_3+2 e_5) V_{cd}^*}{\sqrt{2}}$\\	

$\Lambda_b^0\to   {F'}_{1/2}^{0}  \eta_q  $ &$ \frac{(2 e_1-3 e_2-2 e_3 -e_4-4 e_5) V_{cs}^*}{\sqrt{6}}$&
$\Xi_b^-\to   F_{-1/2}^{--}  K^+  $ & $ (e_1-e_2+e_3+e_4) V_{cs}^*$&
$\Xi_b^0\to   F_{1/2}^{0}  \eta_q  $ & $ \frac{(e_1-e_3-2 e_4+2 e_5) V_{cd}^*}{\sqrt{6}}$\\

&&
$\Xi_b^-\to   F_{-1/2}^-  \overline K^0  $ & $ (e_2+e_4) V_{cs}^*$&
&\\

&&
$\Xi_b^-\to   F_0^-  \pi^0  $ &$ \frac{(e_1-e_2+e_3+e_4+2 e_5) V_{cs}^*}{\sqrt{2}} $&
&\\
&&
$\Xi_b^-\to   F_0^-  \eta_q  $ &$\frac{ (e_1-3 e_2+e_3-e_4+2 e_5) V_{cs}^*}{\sqrt{6}}$&&\\\hline

%%%%%%%%%%%%%%%%%%%%%%%%%%%%%%%%%%%%%%%%%
%%%%%%%%%%%%%%%%%%%%%%%%%%%%%%%%%%%%%%%%%%
$\Sigma_{b}^{+}\to   F_1^{+}   D^0 $ & $ \bar{b}_1 V_{cd}^*$&
$\Omega_{b}^{-}\to   {F'}_0^-   D^0 $ & $ -\bar{b}_2 V_{cd}^*$&
$\Sigma_{b}^{-}\to   F_{-1}^-   D^0 $ & $ (\bar{b}_1+\bar{b}_2) V_{cd}^*$\\

$\Sigma_{b}^{+}\to   F_0^{0}   D^+ $ & $ -\bar{b}_2 V_{cd}^*$&
$\Omega_{b}^{-}\to   F_{-1}^{--}   D^+ $ & $ \bar{b}_2 V_{cd}^*$&
$\Sigma_{b}^{-}\to   F_{-3/2}^{--}   D^+_s $ & $ -(\bar{b}_1+\bar{b}_2) V_{cd}^*$\\

$\Sigma_{b}^{+}\to   F_{3/2}^{+}   D^0 $ & $ -\bar{b}_1 V_{cs}^*$&
$\Omega_{b}^{-}\to   F_{-1/2}^{--}   D^+_s $ & $ -(\bar{b}_1+\bar{b}_2) V_{cd}^*$&
$\Sigma_{b}^{-}\to   F_{-3/2}^{--}   D^+ $ & $ \bar{b}_1 V_{cs}^*$\\

$\Sigma_{b}^{+}\to   F_1^{0}   D^+_s $ & $ \bar{b}_2 V_{cs}^*$&
$\Omega_{b}^{-}\to   F_{1/2}^-   D^0 $ & $ -(\bar{b}_1+\bar{b}_2) V_{cs}^*$&
$\Sigma_{b}^{-}\to   {F'}_{-1/2}^-   D^0 $ & $ \bar{b}_2 V_{cs}^*$\\

$\Sigma_{b}^{+}\to   {F'}_{1/2}^{0}   D^+ $ & $ -\bar{b}_2 V_{cs}^*$&
$\Omega_{b}^{-}\to   F_{-1/2}^{--}   D^+ $ & $ (\bar{b}_1+\bar{b}_2) V_{cs}^*$&
$\Sigma_{b}^{-}\to   F_{-1}^{--}   D^+_s $ & $ -\bar{b}_2 V_{cs}^*$\\

$\Sigma_{b}^{+}\to   F_{1/2}^{0}   D^+_s $ & $ \bar{b}_2 V_{cd}^*$&
$\Omega_{b}^{-}\to   F_{-1/2}^-  \overline K^0  $ & $ -(2 \bar{e}_7+\bar{e}_8) V_{cd}^*$&
$\Sigma_{b}^{-}\to   F_{1/2}^{0}  \pi^-  $ & $ -(\bar{e}_1+\bar{e}_2-\bar{e}_5) V_{cs}^*$
\\

%%%%%%%%%%%%%%%%%%%%%%%%%%%%%%%%%%%%%%%%%%%%%%%%
$\Sigma_{b}^{+}\to   F_{1/2}^{0}  \pi^+  $ & $ -(\bar{e}_2+\bar{e}_6+\bar{e}_8) V_{cs}^*$&
$\Omega_{b}^{-}\to   F_{1/2}^{0}  K^-  $ & $ (2 \bar{e}_7+\bar{e}_8) V_{cd}^*$&
$\Sigma_{b}^{-}\to   F_0^{0}  K^-  $ & $ -(\bar{e}_1+\bar{e}_6) V_{cs}^*$\\

$\Sigma_{b}^{+}\to   {F'}_{1/2}^{0}  \pi^+  $ & $ (\bar{e}_5+2 \bar{e}_7) V_{cs}^*$&
$\Omega_{b}^{-}\to   {F'}_{1/2}^{0}  K^-  $ & $ (\bar{e}_1+\bar{e}_2-\bar{e}_5) V_{cd}^*$&
$\Sigma_{b}^{-}\to   F_{-1}^{--}  K^+  $ & $ (\bar{e}_6-\bar{e}_1) V_{cs}^*$\\

$\Sigma_{b}^{+}\to   F_1^{+}  \overline K^0  $ & $ (\bar{e}_4-\bar{e}_3) V_{cs}^*$&
$\Omega_{b}^{-}\to   F_{-1/2}^{--}  K^+  $ & $ (\bar{e}_1+\bar{e}_3+\bar{e}_4-\bar{e}_6) V_{cd}^*$&
$\Sigma_{b}^{-}\to   {F'}_{-1/2}^-  \pi^0  $ & $ \frac{(\bar{e}_8-\bar{e}_5) V_{cs}^*}{\sqrt{2}}$\\

$\Sigma_{b}^{+}\to   {F'}_{1/2}^{0}  K^+  $ & $ (\bar{e}_2+\bar{e}_6+\bar{e}_8) V_{cd}^*$&
$\Omega_{b}^{-}\to   {F'}_0^-  \overline K^0  $ & $ (-\bar{e}_2+\bar{e}_6+\bar{e}_8) V_{cs}^*$&
$\Sigma_{b}^{-}\to   F_{-1/2}^-  K^0  $ & $ (\bar{e}_5+2 \bar{e}_7) V_{cd}^*$\\

$\Sigma_{b}^{+}\to   F_{3/2}^{+}  K^0  $ & $ (\bar{e}_3-\bar{e}_4) V_{cd}^*$&
$\Omega_{b}^{-}\to   F_0^-  \overline K^0  $ & $ -(\bar{e}_5+2 \bar{e}_7) V_{cs}^*$&
$\Sigma_{b}^{-}\to   {F'}_{-1/2}^-  K^0  $ & $ (\bar{e}_2-\bar{e}_6-\bar{e}_8) V_{cd}^*$\\

$\Sigma_{b}^{+}\to   F_{1/2}^{0}  K^+  $ & $ -(\bar{e}_5+2 \bar{e}_7) V_{cd}^*$&
$\Omega_{b}^{-}\to   F_{1/2}^-  \pi^0  $ & $ \frac{(\bar{e}_3+\bar{e}_4) V_{cs}^*}{\sqrt{2}}$&
$\Sigma_{b}^{-}\to   F_{-3/2}^{--}  K^+  $ & $ (\bar{e}_3+\bar{e}_4) V_{cd}^*$\\

$\Sigma_{b}^{+}\to   F_0^{0}  \pi^+  $ &\tabincell{c}{ $ (\bar{e}_2+\bar{e}_5$\\$+\bar{e}_6+2 \bar{e}_7+\bar{e}_8) V_{cd}^*$}&
$\Omega_{b}^{-}\to   F_1^{0}  K^-  $ &\tabincell{c}{ $ (-\bar{e}_2+\bar{e}_5$\\$+\bar{e}_6+2 \bar{e}_7+\bar{e}_8) V_{cs}^*$}&
$\Sigma_{b}^{-}\to   F_0^{0}  \pi^-  $ &\tabincell{c}{ $ (\bar{e}_2-\bar{e}_5$\\$-\bar{e}_6-2 \bar{e}_7-\bar{e}_8) V_{cd}^*$}\\

$\Sigma_{b}^{+}\to   F_1^{+}  \eta_q  $ &$ \frac{(\bar{e}_2-2 \bar{e}_3-\bar{e}_5+\bar{e}_6+\bar{e}_8) V_{cd}^*}{\sqrt{6}}$&
$\Omega_{b}^{-}\to   F_{1/2}^-  \eta_q  $ & \tabincell{c}{ $ \frac{1}{\sqrt{6}}(2 \bar{e}_2-\bar{e}_3+3 \bar{e}_4+$\\$2 \bar{e}_5-2 \bar{e}_6-2 \bar{e}_8) V_{cs}^*$}&
$\Sigma_{b}^{-}\to   F_{-1}^-  \pi^0  $ &\tabincell{c}{ $ \frac{1}{\sqrt{2}}(-\bar{e}_2  -2 \bar{e}_4$\\$ -\bar{e}_5 +\bar{e}_6 +\bar{e}_8)  V_{\text{cd}}^*$}\\

$\Sigma_{b}^{+}\to   F_1^{+}  \pi^0  $ & $ \frac{(\bar{e}_2-2 \bar{e}_4-\bar{e}_5+\bar{e}_6+\bar{e}_8) V_{cd}^*}{\sqrt{2}}$&
$\Omega_{b}^{-}\to   {F'}_{-1/2}^-  \overline K^0  $ & $ (\bar{e}_1+\bar{e}_2+\bar{e}_5) V_{cd}^*$&
$\Sigma_{b}^{-}\to   F_{-1}^-  \eta_q  $ &\tabincell{c}{ $ \frac{1}{\sqrt{6}}(\bar{e}_2-2 \bar{e}_3$\\$+\bar{e}_5-\bar{e}_6-\bar{e}_8) V_{cd}^*$}\\

$\Sigma_{b}^{+}\to   F_1^{0}  K^+  $ &\tabincell{c}{ $ -(\bar{e}_2+\bar{e}_5$\\$+\bar{e}_6+2 \bar{e}_7+\bar{e}_8) V_{cs}^*$}&
$\Omega_{b}^{-}\to   F_0^-  \eta_q  $ & $ \frac{(\bar{e}_1-2 \bar{e}_5-\bar{e}_6+2 \bar{e}_8) V_{cd}^*}{\sqrt{6}}$&
$\Sigma_{b}^{-}\to   {F'}_{-1/2}^-  \eta_q  $ & $ \frac{(2 \bar{e}_1+\bar{e}_5+2 \bar{e}_6-\bar{e}_8) V_{cs}^*}{\sqrt{6}}$\\

$\Sigma_{b}^{+}\to   F_{3/2}^{+}  \pi^0  $ &\tabincell{c}{ $ \frac{1}{\sqrt{2}}(-\bar{e}_2+\bar{e}_3+\bar{e}_4$\\$+\bar{e}_5-\bar{e}_6-\bar{e}_8) V_{cs}^*$}&
$\Omega_{b}^{-}\to   F_{1/2}^-  K^0  $ & $ (\bar{e}_1+\bar{e}_3-\bar{e}_4+\bar{e}_6) V_{cd}^*$&
$\Sigma_{b}^{-}\to   {F'}_{1/2}^{0}  \pi^-  $ & $ -(2 \bar{e}_7+\bar{e}_8) V_{cs}^*$\\

$\Sigma_{b}^{+}\to   F_{3/2}^{+}  \eta_q  $ &$ \frac{(3 \bar{e}_4-\bar{e}_2-\bar{e}_3++\bar{e}_5-\bar{e}_6-\bar{e}_8) V_{cs}^*}{\sqrt{6}}$&
$\Omega_{b}^{-}\to   F_0^-  \pi^0  $ & $ \frac{(\bar{e}_1-\bar{e}_6) V_{cd}^*}{\sqrt{2}}$&
$\Sigma_{b}^{-}\to   F_0^-  K^0  $ & $ (2 \bar{e}_7+\bar{e}_8) V_{cs}^*$\\

&&
$\Omega_{b}^{-}\to   F_{-1}^{--}  \pi^+  $ & $ (\bar{e}_1-\bar{e}_6) V_{cd}^*$&
$\Sigma_{b}^{-}\to   F_{-1/2}^-  \pi^0  $ & $ \frac{(-\bar{e}_1+\bar{e}_5+\bar{e}_6-\bar{e}_8) V_{cs}^*}{\sqrt{2}}$\\

&&
$\Omega_{b}^{-}\to   F_1^{0}  \pi^-  $ & $ (\bar{e}_1+\bar{e}_6) V_{cd}^*$&
$\Sigma_{b}^{-}\to   F_{-1/2}^-  \eta_q  $ & $ \frac{(-\bar{e}_1-\bar{e}_5+\bar{e}_6+\bar{e}_8) V_{cs}^*}{\sqrt{6}}$\\

&&
$\Omega_{b}^{-}\to   {F'}_0^-  \pi^0  $ & $ -\frac{(\bar{e}_1+\bar{e}_6) V_{cd}^*}{\sqrt{2}}$&
$\Sigma_{b}^{-}\to   F_{-3/2}^{--}  \pi^+  $ & $ -(\bar{e}_1+\bar{e}_3+\bar{e}_4-\bar{e}_6) V_{cs}^*$\\

&&
$\Omega_{b}^{-}\to   {F'}_0^-  \eta_q  $ & $ \frac{(\bar{e}_1+2 \bar{e}_5+\bar{e}_6-2 \bar{e}_8) V_{cd}^*}{\sqrt{6}}$&
$\Sigma_{b}^{-}\to   {F'}_0^-  K^0  $ & $ -(\bar{e}_1+\bar{e}_2+\bar{e}_5) V_{cs}^*$\\
%%%%%%%%%%%%%??4?
&&
$\Omega_{b}^{-}\to   F_{-1/2}^{--}  \pi^+  $ & $ -(\bar{e}_3+\bar{e}_4) V_{cs}^*$&
$\Sigma_{b}^{-}\to   F_{-1}^-  \overline K^0  $ & $ -(\bar{e}_1+\bar{e}_3-\bar{e}_4+\bar{e}_6) V_{cs}^*$\\\hline

\end{tabular}
\end{table}

\begin{table}
\tiny
\caption{The productions of pentaquark $P_{15}$ from singly bottom baryons $T_{b\bar3}/T_{b6}$, with light meson in the final state.}\label{tab:P15_M}
\begin{tabular}{|lr|lr|lr|}\hline\hline
channel & amplitude &channel &amplitude&channel &amplitude \\\hline
%%%%%%%%%%%%%%%%
%%%%%%%%%%%%%%%%%%%%%%%%%%%%%
$\Xi_{b}^{\prime0}\to   F_{1/2}^{0}   D^0 $ & $ -\frac{\bar{b}_2 V_{cd}^*}{\sqrt{2}}$&
$\Xi_{b}^{\prime-}\to   F_{-1/2}^{--}   D^+_s $ & $ -\frac{\bar{b}_2 V_{cs}^*}{\sqrt{2}}$&
$\Sigma_{b}^{0}\to   F_{-1}^-   D^+ $ & $ -\frac{\bar{b}_2 V_{cd}^*}{\sqrt{2}}$\\

$\Xi_{b}^{\prime0}\to   F_{-1/2}^-   D^+ $ & $ \frac{\bar{b}_2 V_{cd}^*}{\sqrt{2}}$&
$\Xi_{b}^{\prime-}\to   {F'}_0^-   D^0 $ & $ -\frac{(2 \bar{b}_1+\bar{b}_2) V_{cs}^*}{\sqrt{2}}$&
$\Sigma_{b}^{0}\to   F_{1/2}^{0}   D^0 $ & $ -\sqrt{2} \bar{b}_1 V_{cs}^*$\\

$\Xi_{b}^{\prime0}\to   {F'}_{-1/2}^-   D^+ $ & $ -\frac{\bar{b}_2 V_{cd}^*}{\sqrt{2}}$&
$\Xi_{b}^{\prime-}\to   F_{-1}^{--}   D^+ $ & $ \frac{(2 \bar{b}_1+\bar{b}_2) V_{cs}^*}{\sqrt{2}}$&
$\Sigma_{b}^{0}\to   F_{-1/2}^-   D^+/D^+_s  $ & $ \sqrt{2}\bar{b}_1 V_{cs}^*/\frac{-(2\bar{b}_1+\bar{b}_2) V_{cd}^*}{\sqrt{2}}$\\

$\Xi_{b}^{\prime0}\to   {F'}_0^-   D^+_s $ & $ \frac{\bar{b}_2 V_{cd}^*}{\sqrt{2}}$&
$\Xi_{b}^{\prime-}\to   F_{-3/2}^{--}   D^+ $ & $ \frac{\bar{b}_2 V_{cd}^*}{\sqrt{2}}$&
$\Sigma_{b}^{0}\to   {F'}_{1/2}^{0}   D^0 $ & $ \frac{\bar{b}_2 V_{cs}^*}{\sqrt{2}}$\\

$\Xi_{b}^{\prime0}\to   {F'}_{1/2}^{0}   D^0 $ & $ \sqrt{2} \bar{b}_1 V_{cd}^*$&
$\Xi_{b}^{\prime-}\to   {F'}_{-1/2}^-   D^0 $ & $ \frac{(2 \bar{b}_1+\bar{b}_2) V_{cd}^*}{\sqrt{2}}$&
$\Sigma_{b}^{0}\to   {F'}_{-1/2}^-   D^+ $ & $ -\frac{\bar{b}_2 V_{cs}^*}{\sqrt{2}}$\\

$\Xi_{b}^{\prime0}\to   F_0^-   D^+_s $ & $ -\sqrt{2} \bar{b}_1 V_{cd}^*$&
$\Xi_{b}^{\prime-}\to   F_{-1}^{--}   D^+_s $ & $ -\frac{(2 \bar{b}_1+\bar{b}_2) V_{cd}^*}{\sqrt{2}}$&
$\Sigma_{b}^{0}\to   {F'}_0^-   D^+_s $ & $ \frac{\bar{b}_2 V_{cs}^*}{\sqrt{2}}$\\

$\Xi_{b}^{\prime0}\to   F_{1/2}^-  D^+_s $ & $\frac{\bar{b}_2 V_{cs}^*}{\sqrt{2}}$&
$\Xi_{b}^{\prime-}\to   F_{1/2}^{0}  K^-  $ & $ \frac{(-\bar{e}_2+\bar{e}_5+\bar{e}_6) V_{cs}^*}{\sqrt{2}}$&
$\Sigma_{b}^{0}\to   F_{0}^-   D^+_s $ & $ -\frac{\bar{b}_2 V_{cs}^*}{\sqrt{2}}$\\

$\Xi_{b}^{\prime0}\to   F_1^{0}   D^0 $ & $ -\frac{(2 \bar{b}_1+\bar{b}_2) V_{cs}^*}{\sqrt{2}}$&
$\Xi_{b}^{\prime-}\to   F_{-1}^{--}  \pi^+  $ & $ \frac{(\bar{e}_6-\bar{e}_1-2 \bar{e}_3-2 \bar{e}_4) V_{cs}^*}{\sqrt{2}}$&
$\Sigma_{b}^{0}\to   F_0^{0}   D^0 $ & $ \frac{(2 \bar{b}_1+\bar{b}_2) V_{cd}^*}{\sqrt{2}}$\\

$\Xi_{b}^{\prime0}\to   F_0^-   D^+ $ & $ \frac{(2 \bar{b}_1+\bar{b}_2) V_{cs}^*}{\sqrt{2}}$&
$\Xi_{b}^{\prime-}\to   F_0^-  \pi^0  $ & $ \frac{(-\bar{e}_1+\bar{e}_5+\bar{e}_6+2 \bar{e}_7) V_{cs}^*}{2}$&	
$\Sigma_{b}^{0}\to   {F'}_{-1/2}^-  K^+  $ & $ \frac{(\bar{e}_2+\bar{e}_6-\bar{e}_8) V_{cd}^*}{\sqrt{2}}$\\

$\Xi_{b}^{\prime0}\to   F_1^{0}  K^0  $ & $ \frac{(\bar{e}_1+2 \bar{e}_3-2 \bar{e}_4+\bar{e}_6) V_{cd}^*}{\sqrt{2}}$&
$\Xi_{b}^{\prime-}\to   {F'}_{1/2}^{0}  \pi^-  $ & $ \frac{(\bar{e}_2-\bar{e}_5-\bar{e}_6) V_{cd}^*}{\sqrt{2}}$&
$\Sigma_{b}^{0}\to   F_0^{0}  \eta_q  $ & $ \frac{(\bar{e}_2-2 \bar{e}_3) V_{cd}^*}{\sqrt{3}}$\\

$\Xi_{b}^{\prime0}\to   {F'}_{1/2}^{0}  \pi^0  $ & $ \frac{(\bar{e}_2-4 \bar{e}_4-\bar{e}_5+\bar{e}_6) V_{cd}^*}{2}$&
$\Xi_{b}^{\prime-}\to   F_{-3/2}^{--}  \pi^+  $ & $ \frac{(\bar{e}_1-\bar{e}_6) V_{cd}^*}{\sqrt{2}}$&
$\Sigma_{b}^{0}\to   {F'}_{1/2}^{0}  K^0  $ & $ \frac{(\bar{e}_2-\bar{e}_6+\bar{e}_8) V_{cd}^*}{\sqrt{2}}$\\

$\Xi_{b}^{\prime0}\to   F_{1/2}^{0}  \pi^0  $ & $ \frac{- (\bar{e}_1+\bar{e}_6-2 \bar{e}_7-\bar{e}_8) V_{cd}^*}{2}$&
$\Xi_{b}^{\prime-}\to   F_{-1}^-  \overline K^0  $ & $ \frac{(\bar{e}_1+\bar{e}_2+\bar{e}_5-\bar{e}_8) V_{cd}^*}{\sqrt{2}}$&
$\Sigma_{b}^{0}\to   F_1^{+}  \pi^-  $ & $ \frac{(\bar{e}_2-\bar{e}_5-\bar{e}_6+\bar{e}_8) V_{cd}^*}{\sqrt{2}}$\\

$\Xi_{b}^{\prime0}\to   F_1^{+}  K^-  $ & $ \frac{(\bar{e}_1+\bar{e}_2-\bar{e}_5+\bar{e}_8) V_{cd}^*}{\sqrt{2}}$&
$\Xi_{b}^{\prime-}\to   F_{-1/2}^-  \pi^0  $ & $ \frac{(\bar{e}_1-\bar{e}_6+2 \bar{e}_7+\bar{e}_8) V_{cd}^*}{2}$&
%$\Sigma_{b}^{0}\to   F_0^{0}  \pi^0  $ & \tabincell{c}{$ -(2 \bar{e}_4+\bar{e}_5$\\$-\bar{e}_6+2 \bar{e}_7+\bar{e}_8) V_{cd}^*$}\\
$\Sigma_{b}^{0}\to   F_0^-  K^+  $ & $ \frac{(\bar{e}_6+2 \bar{e}_7+\bar{e}_8-\bar{e}_1) V_{cs}^*}{\sqrt{2}}$\\

$\Xi_{b}^{\prime0}\to   {F'}_0^-  K^+  $ & $ \frac{(\bar{e}_8-\bar{e}_5) V_{cd}^*}{\sqrt{2}}$&
$\Xi_{b}^{\prime-}\to   F_{-1/2}^{--}  K^+  $ & $ \frac{(\bar{e}_6-\bar{e}_1) V_{cs}^*}{\sqrt{2}}$&
%$\Sigma_{b}^{0}\to   F_{1/2}^{0}  \eta_q  $ &\tabincell{c}{$ -\frac{1}{2\sqrt{3}}(\bar{e}_1+2 \bar{e}_2+2 \bar{e}_3$\\$-6 \bar{e}_4-\bar{e}_5+\bar{e}_6+\bar{e}_8) V_{cs}^*$}\\
$\Sigma_{b}^{0}\to   {F'}_{1/2}^{0}  \eta_q  $ & $ \frac{(2 \bar{e}_1+\bar{e}_5+2 \bar{e}_6-\bar{e}_8) V_{cs}^*}{2 \sqrt{3}}$\\

$\Xi_{b}^{\prime0}\to   {F'}_{-1/2}^-  \pi^+  $ & $ \frac{(\bar{e}_2  +\bar{e}_5  +\bar{e}_6) V_{\text{cd}} {}^*}  {\sqrt{2}}$&
$\Xi_{b}^{\prime-}\to   F_{-1}^{--}  K^+  $ & $ \frac{(\bar{e}_1+2 \bar{e}_3+2 \bar{e}_4-\bar{e}_6) V_{cd}^*}{\sqrt{2}}$&
%$\Sigma_{b}^{0}\to   F_{-1/2}^-  \pi^+  $ &\tabincell{c}{ $ -\frac{1}{\sqrt{2}}(\bar{e}_1+2 \bar{e}_3+2 \bar{e}_4$\\$+\bar{e}_5-\bar{e}_6-\bar{e}_8) V_{cs}^*$}\\
$\Sigma_{b}^{0}\to   F_{-1/2}^-  K^+  $ & $ \frac{(2 \bar{e}_3+2 \bar{e}_4+\bar{e}_5+2 \bar{e}_7) V_{cd}^*}{\sqrt{2}}$\\

$\Xi_{b}^{\prime0}\to   F_{-1/2}^-  \pi^+  $ & $ \frac{(\bar{e}_1-\bar{e}_6-2 \bar{e}_7-\bar{e}_8) V_{cd}^*}{\sqrt{2}}$&
$\Xi_{b}^{\prime-}\to   {F'}_{-1/2}^-  \pi^0  $ & $ \frac{-(\bar{e}_2+4 \bar{e}_4+\bar{e}_5-\bar{e}_6) V_{cd}^*}{2}$&
%$\Sigma_{b}^{0}\to   F_1^{0}  K^0  $ &\tabincell{c}{ $ -\frac{1}{\sqrt{2}}(\bar{e}_1+\bar{e}_2$\\$+\bar{e}_5+2 \bar{e}_7+\bar{e}_8) V_{cs}^*$}\\
$\Sigma_{b}^{0}\to   F_1^{+}  K^-  $ & $ -\frac{(\bar{e}_1+\bar{e}_6) V_{cs}^*}{\sqrt{2}}$\\

$\Xi_{b}^{\prime0}\to   F_{3/2}^{+}  \pi^-  $ & $ \frac{(\bar{e}_1+\bar{e}_6) V_{cd}^*}{\sqrt{2}}$&
$\Xi_{b}^{\prime-}\to   F_{1/2}^{0}  \pi^-  $ & $ \frac{(\bar{e}_1+\bar{e}_6+2 \bar{e}_7+\bar{e}_8) V_{cd}^*}{\sqrt{2}}$&
%$\Sigma_{b}^{0}\to   F_{1/2}^{0}  \pi^0  $ &\tabincell{c}{ $ \frac{1}{2}(\bar{e}_1+2 \bar{e}_3+2 \bar{e}_4$\\$+\bar{e}_5-\bar{e}_6+\bar{e}_8) V_{cs}^*$}\\
$\Sigma_{b}^{0}\to   {F'}_0^-  K^+  $ & $ -\frac{(\bar{e}_2+\bar{e}_5+\bar{e}_6) V_{cs}^*}{\sqrt{2}}$\\

$\Xi_{b}^{\prime0}\to   F_{1/2}^{0}  \overline K^0  $ & $ -\frac{(\bar{e}_2-\bar{e}_6+\bar{e}_8) V_{cs}^*}{\sqrt{2}}$&
$\Xi_{b}^{\prime-}\to   F_0^-  K^0  $ & $ \frac{(\bar{e}_5-\bar{e}_8) V_{cd}^*}{\sqrt{2}}$&
$\Sigma_{b}^{0}\to   F_0^{0}  \overline K^0  $ & $ -\frac{(\bar{e}_1+2 \bar{e}_3-2 \bar{e}_4+\bar{e}_6) V_{cs}^*}{\sqrt{2}}$\\

$\Xi_{b}^{\prime0}\to   {F'}_{1/2}^{0}  \overline K^0  $ & $ \frac{(2 \bar{e}_4+\bar{e}_5+2 \bar{e}_7-2 \bar{e}_3) V_{cs}^*}{\sqrt{2}}$&
$\Xi_{b}^{\prime-}\to   F_{-1/2}^-  \overline K^0  $ & $ \frac{(\bar{e}_8-\bar{e}_5) V_{cs}^*}{\sqrt{2}}$&
$\Sigma_{b}^{0}\to   F_{-1}^-  \pi^+  $ & $ \frac{(\bar{e}_2+\bar{e}_5+\bar{e}_6-\bar{e}_8) V_{cd}^*}{\sqrt{2}}$\\

$\Xi_{b}^{\prime0}\to   F_{3/2}^{+}  K^-  $ & $ \frac{(-\bar{e}_2+\bar{e}_5+\bar{e}_6-\bar{e}_8) V_{cs}^*}{\sqrt{2}}$&
$\Xi_{b}^{\prime-}\to   {F'}_{1/2}^{0}  K^-  $ & $ \frac{-(\bar{e}_1+\bar{e}_6+2 \bar{e}_7+\bar{e}_8) V_{cs}^*}{\sqrt{2}}$&
$\Sigma_{b}^{0}\to   F_{3/2}^{+}  \pi^-  $ & $ -\frac{(\bar{e}_1+\bar{e}_2-\bar{e}_5+\bar{e}_8) V_{cs}^*}{\sqrt{2}}$\\

$\Xi_{b}^{\prime0}\to   F_{1/2}^-  K^+  $ & $ -\frac{(\bar{e}_2+\bar{e}_5+\bar{e}_6-\bar{e}_8) V_{cs}^*}{\sqrt{2}}$&
$\Xi_{b}^{\prime-}\to   F_{1/2}^-  K^0  $ & $ -\frac{(\bar{e}_1+\bar{e}_2+\bar{e}_5-\bar{e}_8) V_{cs}^*}{\sqrt{2}}$&
$\Sigma_{b}^{0}\to   {F'}_{1/2}^{0}  \pi^0  $ & $ \frac{-(\bar{e}_5+4 \bar{e}_7+\bar{e}_8) V_{cs}^*}{2}$\\

$\Xi_{b}^{\prime0}\to   {F'}_0^-  \pi^+  $ & $ -\frac{(\bar{e}_2+\bar{e}_6-\bar{e}_8) V_{cs}^*}{\sqrt{2}}$&
%$\Xi_{b}^{\prime-}\to   {F'}_0^-  \eta_q  $ &\tabincell{c}{ $ \frac{1}{2\sqrt{3}}(\bar{e}_2-2 \bar{e}_3+6 \bar{e}_4$\\$+2 \bar{e}_5-2 \bar{e}_6+\bar{e}_8-\bar{e}_1) V_{cs}^*$}&
$\Xi_{b}^{\prime-}\to   F_0^{0}  K^-  $ & $ \frac{(\bar{e}_1+\bar{e}_2-\bar{e}_5-2 \bar{e}_7-\bar{e}_8) V_{cd}^*}{\sqrt{2}}$&
$\Sigma_{b}^{0}\to   {F'}_{-1/2}^-  \pi^+  $ & $ \frac{(\bar{e}_5-\bar{e}_8) V_{cs}^*}{\sqrt{2}}$\\

$\Xi_{b}^{\prime0}\to   F_0^-  \pi^+  $ & $ \frac{-(2 \bar{e}_3+2 \bar{e}_4+\bar{e}_5+2 \bar{e}_7) V_{cs}^*}{\sqrt{2}}$&
$\Xi_{b}^{\prime-}\to   {F'}_0^-  \pi^0  $ & $ \frac {(\bar{e}_1+\bar{e}_2+2 \bar{e}_3+2 \bar{e}_4-\bar{e}_8) V_{cs}^*}{2}$&
$\Sigma_{b}^{0}\to   F_{1/2}^{0}  K^0  $ & $ \frac{(2 \bar{e}_3-2 \bar{e}_4-\bar{e}_5-2 \bar{e}_7) V_{cd}^*}{\sqrt{2}}$\\

$\Xi_{b}^{\prime0}\to   F_0^-  K^+  $ &\tabincell{c}{ $ \frac{1}{\sqrt{2}}(\bar{e}_1+2 \bar{e}_3+2 \bar{e}_4$\\$+\bar{e}_5-\bar{e}_6-\bar{e}_8) V_{cd}^*$}&
$\Xi_{b}^{\prime-}\to   F_0^-  \eta_q  $ &\tabincell{c}{ $ -\frac{1}{2\sqrt{3}}(\bar{e}_1+\bar{e}_5$\\$-\bar{e}_6+6 \bar{e}_7+2 \bar{e}_8) V_{cs}^*$}&
%$\Sigma_{b}^{0}\to   F_0^-  K^+  $ & $ \frac{(\bar{e}_6+2 \bar{e}_7+\bar{e}_8-\bar{e}_1) V_{cs}^*}{\sqrt{2}}$\\
$\Sigma_{b}^{0}\to   F_0^{0}  \pi^0  $ & \tabincell{c}{$ -(2 \bar{e}_4+\bar{e}_5$\\$-\bar{e}_6+2 \bar{e}_7+\bar{e}_8) V_{cd}^*$}\\

$\Xi_{b}^{\prime0}\to   {F'}_{1/2}^{0}  \eta_q  $ &\tabincell{c}{ $ -\frac{1}{2\sqrt{3}}(2 \bar{e}_1+\bar{e}_2+4 \bar{e}_3$\\$+\bar{e}_5-\bar{e}_6+2 \bar{e}_8) V_{cd}^*$}&
%$\Xi_{b}^{\prime-}\to   F_0^{0}  K^-  $ &\tabincell{c}{ $ \frac{1}{\sqrt{2}}(\bar{e}_1+\bar{e}_2$\\$-\bar{e}_5-2 \bar{e}_7-\bar{e}_8) V_{cd}^*$}&
$\Xi_{b}^{\prime-}\to   {F'}_0^-  \eta_q  $ &\tabincell{c}{ $ \frac{1}{2\sqrt{3}}(\bar{e}_2-2 \bar{e}_3+6 \bar{e}_4$\\$+2 \bar{e}_5-2 \bar{e}_6+\bar{e}_8-\bar{e}_1) V_{cs}^*$}&
%$\Sigma_{b}^{0}\to   {F'}_{1/2}^{0}  \eta_q  $ & $ \frac{(2 \bar{e}_1+\bar{e}_5+2 \bar{e}_6-\bar{e}_8) V_{cs}^*}{2 \sqrt{3}}$\\
$\Sigma_{b}^{0}\to   F_{1/2}^{0}  \eta_q  $ &\tabincell{c}{$ -\frac{1}{2\sqrt{3}}(\bar{e}_1+2 \bar{e}_2+2 \bar{e}_3$\\$-6 \bar{e}_4-\bar{e}_5+\bar{e}_6+\bar{e}_8) V_{cs}^*$}\\

$\Xi_{b}^{\prime0}\to   F_{1/2}^{0}  \eta_q  $ &\tabincell{c}{ $ \frac{1}{2\sqrt{3}}(\bar{e}_1 +2 \bar{e}_5$\\$ +\bar{e}_6 +6 \bar{e}_7  + \bar{e}_8)  V_{\text{cd}} {}^*$}&
$\Xi_{b}^{\prime-}\to   F_{-1/2}^-  \eta_q  $ &\tabincell{c}{ $ \frac{1}{2\sqrt{3}}(\bar{e}_1-2 \bar{e}_5$\\$-\bar{e}_6-6 \bar{e}_7-\bar{e}_8) V_{cd}^*$}&
%$\Sigma_{b}^{0}\to   F_{-1/2}^-  K^+  $ & $ \frac{(2 \bar{e}_3+2 \bar{e}_4+\bar{e}_5+2 \bar{e}_7) V_{cd}^*}{\sqrt{2}}$\\
$\Sigma_{b}^{0}\to   F_{-1/2}^-  \pi^+  $ &\tabincell{c}{ $ -\frac{1}{\sqrt{2}}(\bar{e}_1+2 \bar{e}_3+2 \bar{e}_4$\\$+\bar{e}_5-\bar{e}_6-\bar{e}_8) V_{cs}^*$}\\

$\Xi_{b}^{\prime0}\to   F_0^{0}  \overline K^0  $ &\tabincell{c}{ $ \frac{1}{\sqrt{2}}(\bar{e}_1+\bar{e}_2$\\$+\bar{e}_5+2 \bar{e}_7+\bar{e}_8) V_{cd}^*$}&
$\Xi_{b}^{\prime-}\to   {F'}_{-1/2}^-  \eta_q  $ &\tabincell{c}{ $ -\frac{1}{2\sqrt{3}}(2 \bar{e}_1+\bar{e}_2+4 \bar{e}_3$\\$-\bar{e}_5+\bar{e}_6-2 \bar{e}_8) V_{cd}^*$}&
%$\Sigma_{b}^{0}\to   F_1^{+}  K^-  $ & $ -\frac{(\bar{e}_1+\bar{e}_6) V_{cs}^*}{\sqrt{2}}$\\
$\Sigma_{b}^{0}\to   F_1^{0}  K^0  $ &\tabincell{c}{ $ -\frac{1}{\sqrt{2}}(\bar{e}_1+\bar{e}_2$\\$+\bar{e}_5+2 \bar{e}_7+\bar{e}_8) V_{cs}^*$}\\

$\Xi_{b}^{\prime0}\to   F_1^{0}  \eta_q  $ & \tabincell{c}{$ \frac{1}{2\sqrt{3}}(\bar{e}_2-2 \bar{e}_3+3 (2 \bar{e}_4$\\$+\bar{e}_5-\bar{e}_6+2 \bar{e}_7+\bar{e}_8)) V_{cs}^*$}&
$\Xi_{b}^{\prime-}\to   {F'}_0^-  K^0  $ &\tabincell{c}{ $ \frac{1}{\sqrt{2}}(\bar{e}_1+2 \bar{e}_3-2 \bar{e}_4$\\$-\bar{e}_5+\bar{e}_6+\bar{e}_8) V_{cd}^*$}&
%$\Sigma_{b}^{0}\to   {F'}_0^-  K^+  $ & $ -\frac{(\bar{e}_2+\bar{e}_5+\bar{e}_6) V_{cs}^*}{\sqrt{2}}$\\
$\Sigma_{b}^{0}\to   F_{1/2}^{0}  \pi^0  $ &\tabincell{c}{ $ \frac{1}{2}(\bar{e}_1+2 \bar{e}_3+2 \bar{e}_4$\\$+\bar{e}_5-\bar{e}_6+\bar{e}_8) V_{cs}^*$}\\

$\Xi_{b}^{\prime0}\to   F_1^{0}  \pi^0  $ &\tabincell{c}{ $ \frac{1}{2}(2 \bar{e}_3+2 \bar{e}_4+\bar{e}_5$\\$-\bar{e}_6+2 \bar{e}_7+\bar{e}_8-\bar{e}_2) V_{cs}^*$}&
$\Xi_{b}^{\prime-}\to   {F'}_{-1/2}^-  \overline K^0  $ &\tabincell{c}{ $ -\frac{1}{\sqrt{2}}(\bar{e}_1+2 \bar{e}_3-2 \bar{e}_4$\\$-\bar{e}_5+\bar{e}_6+\bar{e}_8) V_{cs}^*$}&&\\

&&
$\Xi_{b}^{\prime-}\to   F_1^{0}  \pi^-  $ &\tabincell{c}{ $ \frac{1}{\sqrt{2}}(-\bar{e}_1-\bar{e}_2$\\$+\bar{e}_5+2 \bar{e}_7+\bar{e}_8) V_{cs}^*$}&&\\
\hline

%%%%%%%%%%%%%%%%%%%%%%%%%%%%%%%%%%%%%%%%%%%%%%%%%%%%
\end{tabular}
\end{table}
\begin{table}
\tiny
\caption{The productions of pentaquark $P_{15'}$ from singly bottom baryons $T_{b\bar3}/T_{b6}$, with one meson (charmed meson or  light meson) in the final state.}\label{tab:P15p_DM}
\begin{tabular}{|l r|l r|l r|}\hline\hline
channel & amplitude &channel &amplitude &channel &amplitude \\\hline
$\Sigma_{b}^{+}\to   T_{1}^{+}   D^0 $ & $ \bar{c}_1 V_{cd}^*$&
$\Sigma_{b}^{0}\to   T_{0}^{0}   D^0 $ & $ \sqrt{2} \bar{c}_1 V_{cd}^*$&
$\Sigma_{b}^{-}\to   T_{-1}^-   D^0 $ & $ \bar{c}_1 V_{cd}^*$\\
$\Sigma_{b}^{+}\to   T_{0}^{0}   D^+ $ & $ \bar{c}_1 V_{cd}^*$&
$\Sigma_{b}^{0}\to   T_{-1}^-   D^+ $ & $ \sqrt{2} \bar{c}_1 V_{cd}^*$&
$\Sigma_{b}^{-}\to   T_{-2}^{--}   D^+ $ & $ \bar{c}_1 V_{cd}^*$\\
$\Sigma_{b}^{+}\to   T_{3/2}^{+}   D^0 $ & $ \bar{c}_1 V_{cs}^*$&
$\Sigma_{b}^{0}\to   T_{1/2}^{0}   D^0 $ & $ \sqrt{2} \bar{c}_1 V_{cs}^*$&
$\Sigma_{b}^{-}\to   T_{-1/2}^-   D^0 $ & $ \bar{c}_1 V_{cs}^*$\\
$\Sigma_{b}^{+}\to   T_{1/2}^{0}   D^+ $ & $ \bar{c}_1 V_{cs}^*$&
$\Sigma_{b}^{0}\to   T_{-1/2}^-   D^+ $ & $ \sqrt{2} \bar{c}_1 V_{cs}^*$&
$\Sigma_{b}^{-}\to   T_{-3/2}^{--}   D^+ $ & $ \bar{c}_1 V_{cs}^*$\\
$\Sigma_{b}^{+}\to   T_{1/2}^{0}   D^+_s $ & $ \bar{c}_1 V_{cd}^*$&
$\Sigma_{b}^{0}\to   T_{-1/2}^-   D^+_s $ & $ \sqrt{2} \bar{c}_1 V_{cd}^*$&
$\Sigma_{b}^{-}\to   T_{-3/2}^{--}   D^+_s $ & $ \bar{c}_1 V_{cd}^*$\\
$\Sigma_{b}^{+}\to   T_{1}^{0}   D^+_s $ & $ \bar{c}_1 V_{cs}^*$&
$\Sigma_{b}^{0}\to   T_0^-   D^+_s $ & $ \sqrt{2} \bar{c}_1 V_{cs}^*$&
$\Sigma_{b}^{-}\to   T_{-1}^{--}   D^+_s $ & $ \bar{c}_1 V_{cs}^*$\\
%%%%%%%%%%%%%%%%%%%%%%%%%%%%%%%%%
$\Sigma_{b}^{+}\to   T_{1}^{0}  K^+  $ & $ (-\bar{f}_1+\bar{f}_2+2 \bar{f}_3) V_{cs}^*$&
$\Sigma_{b}^{0}\to   T_{1}^{+}  \pi^-  $ & $ \sqrt{2} (\bar{f}_1+\bar{f}_2+\bar{f}_3) V_{cd}^*$&
$\Sigma_{b}^{-}\to   T_{0}^{0}  K^-  $ & $ (\bar{f}_1+\bar{f}_2) V_{cs}^*$\\

$\Sigma_{b}^{+}\to   T_{0}^{0}  \pi^+  $ & $ (-\bar{f}_1+\bar{f}_2+2 \bar{f}_3) V_{cd}^*$&
$\Sigma_{b}^{0}\to   T_{0}^{0}  \pi^0  $ & $ -2 \bar{f}_1 V_{cd}^*$&
$\Sigma_{b}^{-}\to   T_{-1}^-  \pi^0  $ & $ -\sqrt{2} (\bar{f}_1+\bar{f}_3) V_{cd}^*$\\

$\Sigma_{b}^{+}\to   T_{1}^{+}  \eta_q  $ & $ \sqrt{\frac{2}{3}} (\bar{f}_2+\bar{f}_3) V_{cd}^*$&
$\Sigma_{b}^{0}\to   T_{1}^{+}  K^-  $ & $ \sqrt{2} (\bar{f}_1+\bar{f}_2) V_{cs}^*$&
$\Sigma_{b}^{-}\to   T_{-1}^-  \overline K^0  $ & $ (\bar{f}_1+\bar{f}_2) V_{cs}^*$\\

$\Sigma_{b}^{+}\to   T_{1}^{+}  \pi^0  $ & $ \sqrt{2} (\bar{f}_3-\bar{f}_1) V_{cd}^*$&
$\Sigma_{b}^{0}\to   T_{0}^{0}  \overline K^0  $ & $ \sqrt{2} (\bar{f}_1+\bar{f}_2) V_{cs}^*$&
$\Sigma_{b}^{-}\to   T_{-1}^-  \eta_q  $ & $ \sqrt{\frac{2}{3}} (\bar{f}_2+\bar{f}_3) V_{cd}^*$\\

$\Sigma_{b}^{+}\to   T_{1}^{+}  \overline K^0  $ & $ (\bar{f}_1+\bar{f}_2) V_{cs}^*$&
$\Sigma_{b}^{0}\to   T_{0}^{0}  \eta_q  $ & $ \frac{2 (\bar{f}_2+\bar{f}_3) V_{cd}^*}{\sqrt{3}}$&
$\Sigma_{b}^{-}\to   T_{1/2}^{0}  \pi^-  $ & $ 2 \bar{f}_3 V_{cs}^*$\\

$\Sigma_{b}^{+}\to   T_2^{++}  K^-  $ & $ (\bar{f}_1+\bar{f}_2) V_{cs}^*$&
$\Sigma_{b}^{0}\to   T_{-1}^-  \pi^+  $ & $\sqrt{2}(- \bar{f}_1+\bar{f}_2+ \bar{f}_3)  V_{cd}^*$&
$\Sigma_{b}^{-}\to   T_{-2}^{--}  \pi^+  $ & $ (\bar{f}_2-\bar{f}_1) V_{cd}^*$\\

$\Sigma_{b}^{+}\to   T_2^{++}  \pi^-  $ & $ (\bar{f}_1+\bar{f}_2) V_{cd}^*$&
$\Sigma_{b}^{0}\to   T_{3/2}^{+}  \pi^-  $ & $ \sqrt{2} \bar{f}_3 V_{cs}^*$&
$\Sigma_{b}^{-}\to   T_{-1/2}^-  \pi^0  $ & $ \frac{(-\bar{f}_1+\bar{f}_2-2 \bar{f}_3) V_{cs}^*}{\sqrt{2}}$\\

$\Sigma_{b}^{+}\to   T_{1/2}^{0}  K^+  $ & $ (-\bar{f}_1+\bar{f}_2+2 \bar{f}_3) V_{cd}^*$&
$\Sigma_{b}^{0}\to   T_{1/2}^{0}  K^0  $ & $ \sqrt{2} (\bar{f}_1+\bar{f}_2+\bar{f}_3) V_{cd}^*$&
$\Sigma_{b}^{-}\to   T_{-1/2}^-  K^0  $ & $ (\bar{f}_1+\bar{f}_2+2 \bar{f}_3) V_{cd}^*$\\

$\Sigma_{b}^{+}\to   T_{1/2}^{0}  \pi^+  $ & $ (-\bar{f}_1+\bar{f}_2+2 \bar{f}_3) V_{cs}^*$&
$\Sigma_{b}^{0}\to   T_{1/2}^{0}  \pi^0  $ & $ (\bar{f}_2-\bar{f}_1) V_{cs}^*$&
$\Sigma_{b}^{-}\to   T_{-1/2}^-  \eta_q  $ & $ -\frac{(3 \bar{f}_1+\bar{f}_2-2 \bar{f}_3) V_{cs}^*}{\sqrt{6}}$\\

$\Sigma_{b}^{+}\to   T_{3/2}^{+}  K^0  $ & $ (\bar{f}_1+\bar{f}_2) V_{cd}^*$&
$\Sigma_{b}^{0}\to   T_{1/2}^{0}  \eta_q  $ & $ -\frac{(3 \bar{f}_1+\bar{f}_2-2 \bar{f}_3) V_{cs}^*}{\sqrt{3}}$&
$\Sigma_{b}^{-}\to   T_{-3/2}^{--}  \pi^+  $ & $ (\bar{f}_2-\bar{f}_1) V_{cs}^*$\\

$\Sigma_{b}^{+}\to   T_{3/2}^{+}  \eta_q  $ & $ -\frac{(3 \bar{f}_1+\bar{f}_2-2 \bar{f}_3) V_{cs}^*}{\sqrt{6}}$&
$\Sigma_{b}^{0}\to   T_{-1/2}^-  \pi^+  $ & $ \sqrt{2} (-\bar{f}_1+\bar{f}_2+\bar{f}_3) V_{cs}^*$&
$\Sigma_{b}^{-}\to   T_0^-  K^0  $ & $ 2 \bar{f}_3 V_{cs}^*$\\

$\Sigma_{b}^{+}\to   T_{3/2}^{+}  \pi^0  $ & $ \frac{(-\bar{f}_1+\bar{f}_2+2 \bar{f}_3) V_{cs}^*}{\sqrt{2}}$&
$\Sigma_{b}^{0}\to   T_{-1/2}^-  K^+  $ & $ \sqrt{2}(- \bar{f}_1 + \bar{f}_2+ \bar{f}_3)  V_{cd}^*$&
$\Sigma_{b}^{-}\to   T_{-3/2}^{--}  K^+  $ & $ (\bar{f}_2-\bar{f}_1) V_{cd}^*$\\

&&
$\Sigma_{b}^{0}\to   T_{1}^{0}  K^0  $ & $ \sqrt{2} \bar{f}_3 V_{cs}^*$&
$\Sigma_{b}^{-}\to   T_{0}^{0}  \pi^-  $ & $ (\bar{f}_1+\bar{f}_2+2 \bar{f}_3) V_{cd}^*$\\

&&
$\Sigma_{b}^{0}\to   T_0^-  K^+  $ & $ \sqrt{2} (-\bar{f}_1+\bar{f}_2+\bar{f}_3) V_{cs}^*$&
$\Sigma_{b}^{-}\to   T_{-1}^{--}  K^+  $ & $ (\bar{f}_2-\bar{f}_1) V_{cs}^*$\\\hline
%%%%%%%%%%%%%%%%%%%%%%%%%%%%%%%%%
%%%%%%%%%%%%%%%%%%%%%%%%%%%%%%%%
$\Xi_{b}^{\prime0}\to   T_{1/2}^{0}   D^0 $ & $ \sqrt{2} \bar{c}_1 V_{cd}^*$&
$\Xi_{b}^{\prime-}\to   T_{-1/2}^-   D^0 $ & $ \sqrt{2} \bar{c}_1 V_{cd}^*$&
$\Omega_{b}^{-}\to   T_0^-   D^0 $ & $ \bar{c}_1 V_{cd}^*$\\
$\Xi_{b}^{\prime0}\to   T_{-1/2}^-   D^+ $ & $ \sqrt{2} \bar{c}_1 V_{cd}^*$&
$\Xi_{b}^{\prime-}\to   T_{-3/2}^{--}   D^+ $ & $ \sqrt{2} \bar{c}_1 V_{cd}^*$&
$\Omega_{b}^{-}\to   T_{-1}^{--}   D^+ $ & $ \bar{c}_1 V_{cd}^*$\\
$\Xi_{b}^{\prime0}\to   T_{1}^{0}   D^0 $ & $ \sqrt{2} \bar{c}_1 V_{cs}^*$&
$\Xi_{b}^{\prime-}\to   T_0^-   D^0 $ & $ \sqrt{2} \bar{c}_1 V_{cs}^*$&
$\Omega_{b}^{-}\to   T_{1/2}^-   D^0 $ & $ \bar{c}_1 V_{cs}^*$\\
$\Xi_{b}^{\prime0}\to   T_0^-   D^+ $ & $ \sqrt{2} \bar{c}_1 V_{cs}^*$&
$\Xi_{b}^{\prime-}\to   T_{-1}^{--}   D^+ $ & $ \sqrt{2} \bar{c}_1 V_{cs}^*$&
$\Omega_{b}^{-}\to   T_{-1/2}^{--}   D^+ $ & $ \bar{c}_1 V_{cs}^*$\\
$\Xi_{b}^{\prime0}\to   T_0^-   D^+_s $ & $ \sqrt{2} \bar{c}_1 V_{cd}^*$&
$\Xi_{b}^{\prime-}\to   T_{-1}^{--}   D^+_s $ & $ \sqrt{2} \bar{c}_1 V_{cd}^*$&
$\Omega_{b}^{-}\to   T_{-1/2}^{--}   D^+_s $ & $ \bar{c}_1 V_{cd}^*$\\
$\Xi_{b}^{\prime0}\to   T_{1/2}^-   D^+_s $ & $ \sqrt{2} \bar{c}_1 V_{cs}^*$&
$\Xi_{b}^{\prime-}\to   T_{-1/2}^{--}   D^+_s $ & $ \sqrt{2} \bar{c}_1 V_{cs}^*$&
$\Omega_{b}^{-}\to   T_{0}^{--}   D^+_s $ & $ \bar{c}_1 V_{cs}^*$\\
%%%%%%%%%%%%%%%%%%%%%%%%%%%%%%%%%%%%%%%%%%%%%%%
$\Xi_{b}^{\prime0}\to   T_{1}^{+}  K^-  $ & $ \sqrt{2} \bar{f}_3 V_{cd}^*$&
$\Xi_{b}^{\prime-}\to   T_{0}^{0}  K^-  $ & $ \sqrt{2} \bar{f}_3 V_{cd}^*$&
$\Omega_{b}^{-}\to   T_{1/2}^{0}  K^-  $ & $ 2 \bar{f}_3 V_{cd}^*$\\

$\Xi_{b}^{\prime0}\to   T_{0}^{0}  \overline K^0  $ & $ \sqrt{2} \bar{f}_3 V_{cd}^*$&
$\Xi_{b}^{\prime-}\to   T_{1/2}^{0}  \pi^-  $ & $ \sqrt{2} (\bar{f}_1+\bar{f}_2+\bar{f}_3) V_{cd}^*$&
$\Omega_{b}^{-}\to   T_{-1/2}^-  \overline K^0  $ & $ 2 \bar{f}_3 V_{cd}^*$\\

$\Xi_{b}^{\prime0}\to   T_{3/2}^{+}  \pi^-  $ & $ \sqrt{2} (\bar{f}_1+\bar{f}_2) V_{cd}^*$&
$\Xi_{b}^{\prime-}\to   T_{-1}^-  \overline K^0  $ & $ \sqrt{2} \bar{f}_3 V_{cd}^*$&
$\Omega_{b}^{-}\to   T_{1}^{0}  \pi^-  $ & $ (\bar{f}_1+\bar{f}_2) V_{cd}^*$\\

$\Xi_{b}^{\prime0}\to   T_{1/2}^{0}  \pi^0  $ & $ (\bar{f}_3-2 \bar{f}_1) V_{cd}^*$&
$\Xi_{b}^{\prime-}\to   T_{1/2}^{0}  K^-  $ & $ \sqrt{2} (\bar{f}_1+\bar{f}_2+\bar{f}_3) V_{cs}^*$&
$\Omega_{b}^{-}\to   T_0^-  \pi^0  $ & $ -\sqrt{2} \bar{f}_1 V_{cd}^*$\\

$\Xi_{b}^{\prime0}\to   T_{3/2}^{+}  K^-  $ & $ \sqrt{2} (\bar{f}_1+\bar{f}_2+\bar{f}_3) V_{cs}^*$&
$\Xi_{b}^{\prime-}\to   T_{-1/2}^-  \pi^0  $ & $ -(2 \bar{f}_1+\bar{f}_3) V_{cd}^*$&
$\Omega_{b}^{-}\to   T_{1}^{0}  K^-  $ & $ (\bar{f}_1+\bar{f}_2+2 \bar{f}_3) V_{cs}^*$\\

$\Xi_{b}^{\prime0}\to   T_{1/2}^{0}  \overline K^0  $ & $ \sqrt{2} (\bar{f}_1+\bar{f}_2+\bar{f}_3) V_{cs}^*$&
$\Xi_{b}^{\prime-}\to   T_{-1/2}^-  \overline K^0  $ & $ \sqrt{2} (\bar{f}_1+\bar{f}_2+\bar{f}_3) V_{cs}^*$&
$\Omega_{b}^{-}\to   T_0^-  \overline K^0  $ & $ (\bar{f}_1+\bar{f}_2+2 \bar{f}_3) V_{cs}^*$\\

$\Xi_{b}^{\prime0}\to   T_{1/2}^{0}  \eta_q  $ & $ \frac{(2 \bar{f}_2-\bar{f}_3) V_{cd}^*}{\sqrt{3}}$&
$\Xi_{b}^{\prime-}\to   T_{-1/2}^-  \eta_q  $ & $ \frac{(2 \bar{f}_2-\bar{f}_3) V_{cd}^*}{\sqrt{3}}$&
$\Omega_{b}^{-}\to   T_0^-  \eta_q  $ & $ \sqrt{\frac{2}{3}} (\bar{f}_2-2 \bar{f}_3) V_{cd}^*$\\

$\Xi_{b}^{\prime0}\to   T_{-1/2}^-  \pi^+  $ & $ \sqrt{2} (-\bar{f}_1+ \bar{f}_2+ \bar{f}_3)  V_{cd}^*$&
$\Xi_{b}^{\prime-}\to   T_{1}^{0}  \pi^-  $ & $ \sqrt{2} \bar{f}_3 V_{cs}^*$&
$\Omega_{b}^{-}\to   T_{-1}^{--}  \pi^+  $ & $ (\bar{f}_2-\bar{f}_1) V_{cd}^*$\\

$\Xi_{b}^{\prime0}\to   T_{1}^{0}  \pi^0  $ & $ (-\bar{f}_1+\bar{f}_2+\bar{f}_3) V_{cs}^*$&
$\Xi_{b}^{\prime-}\to   T_{-3/2}^{--}  \pi^+  $ & $ \sqrt{2} (\bar{f}_2-\bar{f}_1) V_{cd}^*$&
$\Omega_{b}^{-}\to   T_{1/2}^-  \pi^0  $ & $ \frac{(\bar{f}_2-\bar{f}_1) V_{cs}^*}{\sqrt{2}}$\\

$\Xi_{b}^{\prime0}\to   T_{1}^{0}  \eta_q  $ & $ -\frac{(3 \bar{f}_1+\bar{f}_2+\bar{f}_3) V_{cs}^*}{\sqrt{3}}$&
$\Xi_{b}^{\prime-}\to   T_0^-  \pi^0  $ & $ -(\bar{f}_1-\bar{f}_2+\bar{f}_3) V_{cs}^*$&
$\Omega_{b}^{-}\to   T_{1/2}^-  \eta_q  $ & $ -\frac{(3 \bar{f}_1+\bar{f}_2+4 \bar{f}_3) V_{cs}^*}{\sqrt{6}}$\\

$\Xi_{b}^{\prime0}\to   T_{1}^{0}  K^0  $ & $ \sqrt{2} (\bar{f}_1+\bar{f}_2) V_{cd}^*$&
$\Xi_{b}^{\prime-}\to   T_0^-  K^0  $ & $ \sqrt{2} (\bar{f}_1+\bar{f}_2+\bar{f}_3) V_{cd}^*$&
$\Omega_{b}^{-}\to   T_{1/2}^-  K^0  $ & $ (\bar{f}_1+\bar{f}_2) V_{cd}^*$\\

$\Xi_{b}^{\prime0}\to   T_0^-  \pi^+  $ & $ \sqrt{2} (-\bar{f}_1+\bar{f}_2+\bar{f}_3) V_{cs}^*$&
$\Xi_{b}^{\prime-}\to   T_0^-  \eta_q  $ & $ -\frac{(3 \bar{f}_1+\bar{f}_2+\bar{f}_3) V_{cs}^*}{\sqrt{3}}$&
$\Omega_{b}^{-}\to   T_{-1/2}^{--}  \pi^+  $ & $ (\bar{f}_2-\bar{f}_1) V_{cs}^*$\\

$\Xi_{b}^{\prime0}\to   T_0^-  K^+  $ & $ \sqrt{2}(- \bar{f}_1+ \bar{f}_2+ \bar{f}_3) V_{cd}^*$&
$\Xi_{b}^{\prime-}\to   T_{-1}^{--}  \pi^+  $ & $ \sqrt{2} (\bar{f}_2-\bar{f}_1) V_{cs}^*$&
$\Omega_{b}^{-}\to   T_{-1/2}^{--}  K^+  $ & $ (\bar{f}_2-\bar{f}_1) V_{cd}^*$\\

$\Xi_{b}^{\prime0}\to   T_{1/2}^-  K^+  $ & $ \sqrt{2} (-\bar{f}_1+\bar{f}_2+\bar{f}_3) V_{cs}^*$&
$\Xi_{b}^{\prime-}\to   T_{1/2}^-  K^0  $ & $ \sqrt{2} \bar{f}_3 V_{cs}^*$&
$\Omega_{b}^{-}\to   T_{0}^{--}  K^+  $ & $ (\bar{f}_2-\bar{f}_1) V_{cs}^*$\\
&&
$\Xi_{b}^{\prime-}\to   T_{-1}^{--}  K^+  $ & $ \sqrt{2} (\bar{f}_2-\bar{f}_1) V_{cd}^*$&&\\
&&
$\Xi_{b}^{\prime-}\to   T_{-1/2}^{--}  K^+  $ & $ \sqrt{2} (\bar{f}_2-\bar{f}_1) V_{cs}^*$&&\\\hline
%%%%%%%%%%%%%%%%%%%%%%%%%%%%%%%%%%%
%%%%%%%%%%%%%%%%%%%%%%%%%%%%%%%%
$\Xi_b^0\to   T_{1}^{+}  K^-  $ & $ 2 f_2 V_{cd}^*$&
$\Xi_b^-\to   T_{0}^{0}  K^-  $ & $ 2 f_2 V_{cd}^*$&
$\Lambda_b^0\to   T_{1}^{+}  \pi^-  $ & $ 2 f_2 V_{cd}^*$\\
$\Xi_b^0\to   T_{0}^{0}  \overline K^0  $ & $ 2 f_2 V_{cd}^*$&
$\Xi_b^-\to   T_{-1}^-  \overline K^0  $ & $ 2 f_2 V_{cd}^*$&
$\Lambda_b^0\to   T_{0}^{0}  \pi^0  $ & $ -2 \sqrt{2} f_2 V_{cd}^*$\\
$\Xi_b^0\to   T_{3/2}^{+}  K^-  $ & $ 2 f_2 V_{cs}^*$&
$\Xi_b^-\to   T_{1/2}^{0}  \pi^-  $ & $ -2 f_2 V_{cd}^*$&
$\Lambda_b^0\to   T_{-1}^-  \pi^+  $ & $ -2 f_2 V_{cd}^*$\\
$\Xi_b^0\to   T_{1/2}^{0}  \pi^0  $ & $ -\sqrt{2} f_2 V_{cd}^*$&
$\Xi_b^-\to   T_{1/2}^{0}  K^-  $ & $ 2 f_2 V_{cs}^*$&
$\Lambda_b^0\to   T_{3/2}^{+}  \pi^-  $ & $ 2 f_2 V_{cs}^*$\\
$\Xi_b^0\to   T_{1/2}^{0}  \overline K^0  $ & $ 2 f_2 V_{cs}^*$&
$\Xi_b^-\to   T_{-1/2}^-  \pi^0  $ & $ \sqrt{2} f_2 V_{cd}^*$&
$\Lambda_b^0\to   T_{1/2}^{0}  \pi^0  $ & $ -2 \sqrt{2} f_2 V_{cs}^*$\\
$\Xi_b^0\to   T_{1/2}^{0}  \eta_q  $ & $ -\sqrt{6} f_2 V_{cd}^*$&
$\Xi_b^-\to   T_{-1/2}^-  \overline K^0  $ & $ 2 f_2 V_{cs}^*$&
$\Lambda_b^0\to   T_{1/2}^{0}  K^0  $ & $ 2 f_2 V_{cd}^*$\\
$\Xi_b^0\to   T_{-1/2}^-  \pi^+  $ & $ -2 f_2 V_{cd}^*$&
$\Xi_b^-\to   T_{-1/2}^-  \eta_q  $ & $ -\sqrt{6} f_2 V_{cd}^*$&
$\Lambda_b^0\to   T_{-1/2}^-  \pi^+  $ & $ -2 f_2 V_{cs}^*$\\
$\Xi_b^0\to   T_{1}^{0}  \pi^0  $ & $ -\sqrt{2} f_2 V_{cs}^*$&
$\Xi_b^-\to   T_{1}^{0}  \pi^-  $ & $ -2 f_2 V_{cs}^*$&
$\Lambda_b^0\to   T_{-1/2}^-  K^+  $ & $ -2 f_2 V_{cd}^*$\\
$\Xi_b^0\to   T_{1}^{0}  \eta_q  $ & $ -\sqrt{6} f_2 V_{cs}^*$&
$\Xi_b^-\to   T_0^-  \pi^0  $ & $ \sqrt{2} f_2 V_{cs}^*$&
$\Lambda_b^0\to   T_{1}^{0}  K^0  $ & $ 2 f_2 V_{cs}^*$\\
$\Xi_b^0\to   T_0^-  \pi^+  $ & $ -2 f_2 V_{cs}^*$&
$\Xi_b^-\to   T_0^-  K^0  $ & $ -2 f_2 V_{cd}^*$&
$\Lambda_b^0\to   T_0^-  K^+  $ & $ -2 f_2 V_{cs}^*$\\
$\Xi_b^0\to   T_0^-  K^+  $ & $ -2 f_2 V_{cd}^*$&
$\Xi_b^-\to   T_0^-  \eta_q  $ & $ -\sqrt{6} f_2 V_{cs}^*$&&\\
$\Xi_b^0\to   T_{1/2}^-  K^+  $ & $ -2 f_2 V_{cs}^*$&
$\Xi_b^-\to   T_{1/2}^-  K^0  $ & $ -2 f_2 V_{cs}^*$&&\\\hline
%%%%%%%%%%%%%%%%%%%%%%%%%%%%%%%%%%%%%%%%%%%
%%%%%%%%%%%%%%%%%%%%%%%%%%%%%%%%%%%%%%%%%%%%

%%%%%%%%%%%%%%%%%%%%%%%%%%%%%%%%%%%%%%%%%%%%%%%%%%

%%%%%%%%%%%%%%%%%%%%%%%%%%%%%%%%%
\end{tabular}
\end{table}

\section{tensor forms}
\label{sec:app}
The pentaquark state $\bar c qqqq$($T^{ijkl}$) can be decomposed into different tensor representations upon the SU(3) flavor symmetry.
\begin{eqnarray}
T^{ijkl}&:&3 \otimes 3 \otimes 3 \otimes 3\\\nonumber
&=& 6 \otimes 6 + 6 \otimes \bar{3} + \bar{3} \otimes 6 + \bar{3} \otimes \bar{3}\\\nonumber
&=&3\oplus \bar{3} \oplus \bar{3} \oplus \bar{6} \oplus \bar{6} \oplus 15 \oplus 15 \oplus 15 \oplus 15'.\nonumber
\end{eqnarray}
The tensor reduction Eq.~\eqref{eq:tensor} can tell the representations of irreducible tensors $\hat{T_3}$, $\Bar{T}_{\bar 6}$, $\widetilde{T}_{15}, \ldots$, with the way of projection. In addition, the traceless tensor $15$ indicates the following equations.
\begin{eqnarray*}
T_1^{\{11\}}=-T_2^{\{12\}}-T_3^{\{13\}},\\
T_2^{\{22\}}=-T_1^{\{21\}}-T_3^{\{23\}},\\
T_3^{\{33\}}=-T_1^{\{31\}}-T_{2}^{\{32\}}.
\end{eqnarray*}
Here, filtering out redundant mathematical calculations, we directly list the tensor representations $\bar 6$, $15$ and $15'$ respectively. The $\bar6$ states can be obtained from $\bar3\otimes \bar3\to T_{\bar6}$ or $6\otimes 6\to T'_{\bar6}$,
\begin{eqnarray*}
(T_{\bar6})_{\{11\}} &=& \bar{b}dsds-\bar{c}dssd-\bar{c}sdds+\bar{c}sdsd, \quad
(T_{\bar6})_{\{12\}}  = \bar{c}dssu-\bar{c}dsus-\bar{c}sdsu+\bar{c}sdus, \\
(T_{\bar6})_{\{13\}}  &=&  -\bar{c}dsdu+\bar{c}dsud+\bar{c}sddu-\bar{c}sdud, \quad
 %$3\otimes 3$ & $(T_{\BAR6})_{\{21\}}$ & $\bar{c}suds}-\bar{c}susd}-\bar{c}usds}+\bar{c}ussd}$ \\
(T_{\bar6})_{\{22\}}  =  \bar{c}susu-\bar{c}suus-\bar{c}ussu+\bar{c}usus, \\
(T_{\bar6})_{\{23\}}  &=&  -\bar{c}sudu+\bar{c}suud+\bar{c}usdu-\bar{c}usud, \quad
 %$3\otimes 3$ & $(T_6)_{\{31\}}$ & $-\bar{c}duds}+\bar{c}dusd}+\bar{c}udds}-\bar{c}udsd}$ \\
 %$3\otimes 3$ & $(T_6)_{\{32\}}$ & $-\bar{c}dusu}+\bar{c}duus}+\bar{c}udsu}-\bar{c}udus}$ \\
(T_{\bar6})_{\{33\}} =  \bar{c}dudu-\bar{c}duud-\bar{c}uddu+\bar{c}udud,\\
%6\otimes 6&&\\
(T'_{\bar6})_{\{11\}}  &=&  \bar{c}ddss+\frac{1}{2} (-\bar{c}dsds-\bar{c}dssd-\bar{c}sdds-\bar{c}sdsd)+\bar{c}ssdd, \\
(T'_{\bar6})_{\{12\}}  &=&  \frac{1}{4} (\bar{c}dssu+\bar{c}dsus-2 \bar{c}duss+\bar{c}sdsu+\bar{c}sdus-2 \bar{c}ssdu-2 \bar{c}ssud+\bar{c}suds\\&&+\bar{c}susd-2 \bar{c}udss+\bar{c}usds+\bar{c}ussd), \\
(T'_{\bar6})_{\{13\}}  &=&  \frac{1}{4} (-2 \bar{c}ddsu-2 \bar{c}ddus+\bar{c}dsdu+\bar{c}dsud+\bar{c}duds+\bar{c}dusd+\bar{c}sddu+\bar{c}sdud\\&&-2 \bar{c}sudd+\bar{c}udds+\bar{c}udsd-2 \bar{c}usdd), \\
 %$6\otimes 6$ & $(T_{\BAR6})_{\{21\}}$ & $\frac{1}{4} (\bar{c}dssu+\bar{c}dsus-2 \bar{c}duss}+\bar{c}sdsu+\bar{c}sdus-2 \bar{c}ssdu}-2 \bar{c}ssud}+\bar{c}suds}+\bar{c}susd}-2 \bar{c}udss}+\bar{c}usds}+\bar{c}ussd})$ \\
(T'_{\bar6})_{\{22\}}  &=&  \bar{c}ssuu+\frac{1}{2} (-\bar{c}susu-\bar{c}suus-\bar{c}ussu-\bar{c}usus)+\bar{c}uuss,\\
(T'_{\bar6})_{\{23\}}  &=&  \frac{1}{4} (-2 \bar{c}dsuu+\bar{c}dusu+\bar{c}duus-2 \bar{c}sduu+\bar{c}sudu+\bar{c}suud+\bar{c}udsu+\bar{c}udus\\&&+\bar{c}usdu+\bar{c}usud-2 \bar{c}uuds-2 \bar{c}uusd), \\
 %$6\otimes 6$ & $(T_{\BAR6})_{\{31\}}$ & $\frac{1}{4} (-2 \bar{c}ddsu}-2 \bar{c}ddus}+\bar{c}dsdu}+\bar{c}dsud}+\bar{c}duds}+\bar{c}dusd}+\bar{c}sddu}+\bar{c}sdud}-2 \bar{c}sudd}+\bar{c}udds}+\bar{c}udsd}-2 \bar{c}usdd})$ \\
 %$6\otimes 6$ & $(T_{\BAR6})_{\{32\}}$ & $\frac{1}{4} (-2 \bar{c}dsuu}+\bar{c}dusu}+\bar{c}duus}-2 \bar{c}sduu}+\bar{c}sudu+\bar{c}suud+\bar{c}udsu}+\bar{c}udus}+\bar{c}usdu+\bar{c}usud}-2 \bar{c}uuds}-2 \bar{c}uusd})$ \\
(T'_{\bar6})_{\{33\}}  &=&  \bar{c}dduu+\frac{1}{2} (-\bar{c}dudu-\bar{c}duud-\bar{c}uddu-\bar{c}udud)+\bar{c}uudd.
\end{eqnarray*}
The 15 state can be decomposed from $\bar3\otimes 6\to T_{15}$, $6\otimes \bar3\to T'_{15}$ or $6\otimes 6\to T''_{15}$. Easily, the $T'_{15}$ can be obtained by exchanging two pair light quarks $[q_1 q_2]\leftrightarrow [q_3 q_4]$ in $T_{15} \ (\bar b [q_1 q_2][q_3 q_4])$, such as $ (T'_{15})^{\{11\}}_2=\bar{c}uusu-\bar{c}uuus$.
\begin{small}
\begin{eqnarray*}
(T_{15})^{\{11\}}_1&=&\frac{1}{4} (2 \bar{c}dsuu+\bar{c}dusu+\bar{c}duus-2 \bar{c}sduu-\bar{c}sudu-\bar{c}suud-\bar{c}udsu-\bar{c}udus+\bar{c}usdu+\bar{c}usud), \\
(T_{15})^{\{11\}}_2&=&\bar{c}suuu-\bar{c}usuu, \quad
(T_{15})^{\{11\}}_3=\bar{c}uduu-\bar{c}duuu, \quad
(T_{15})^{\{22\}}_1=\bar{c}dsdd-\bar{c}sddd, \\
(T_{15})^{\{22\}}_3&=&\bar{c}uddd-\bar{c}dudd, \quad
(T_{15})^{\{33\}}_1=\bar{c}dsss-\bar{c}sdss, \quad
(T_{15})^{\{33\}}_2=\bar{c}suss-\bar{c}usss, \\
(T_{15})^{\{12\}}_1&=&\frac{3}{8}( \bar{c}dsdu+ \bar{c}dsud- \bar{c}sddu- \bar{c}sdud)+\frac{1}{8}(\bar{c}duds+\bar{c}dusd-\bar{c}udds-\bar{c}udsd)+\frac{2}{8}( \bar{c}usdd- \bar{c}sudd), \\
(T_{15})^{\{12\}}_2&=&\frac{1}{8} (\bar{c}dusu-2 \bar{c}dsuu+\bar{c}duus+2 \bar{c}sduu+3 \bar{c}sudu+3 \bar{c}suud-\bar{c}udsu-\bar{c}udus-3 \bar{c}usdu-3 \bar{c}usud), \\
(T_{15})^{\{12\}}_3&=&\frac{1}{2} (-\bar{c}dudu-\bar{c}duud+\bar{c}uddu+\bar{c}udud),\ (T_{15})^{\{13\}}_2=\frac{1}{2} (\bar{c}susu+\bar{c}suus-\bar{c}ussu-\bar{c}usus), \\
(T_{15})^{\{13\}}_1&=&\frac{1}{8} (3 \bar{c}dssu+3 \bar{c}dsus+2 \bar{c}duss-3 \bar{c}sdsu-3 \bar{c}sdus-\bar{c}suds-\bar{c}susd-2 \bar{c}udss+\bar{c}usds+\bar{c}ussd), \\
(T_{15})^{\{13\}}_3&=&\frac{1}{8} (3 \bar{c}udus-2 \bar{c}dsuu-3 \bar{c}dusu-3 \bar{c}duus+2 \bar{c}sduu-\bar{c}sudu-\bar{c}suud+3 \bar{c}udsu+\bar{c}usdu+\bar{c}usud), \\
%(T_{15})^{\{21\}}_1&=&\frac{1}{8} (3 \bar{c}dsdu+3 \bar{c}dsud+\bar{c}duds+\bar{c}dusd-3 \bar{c}sddu-3 \bar{c}sdud-2 \bar{c}sudd-\bar{c}udds-\bar{c}udsd+2 \bar{c}usdd) \\
%(T_{15})^{\{21\}}_2&=&\frac{1}{8} (-2 \bar{c}dsuu+\bar{c}dusu+\bar{c}duus+2 \bar{c}sduu+3 \bar{c}sudu+3 \bar{c}suud-\bar{c}udsu-\bar{c}udus-3 \bar{c}usdu-3 \bar{c}usud) \\
%(T_{15})^{\{21\}}_3&=&\frac{1}{2} (-\bar{c}dudu-\bar{c}duud+\bar{c}uddu+\bar{c}udud) \\
(T_{15})^{\{22\}}_2&=&\frac{1}{4} (-\bar{c}dsdu-\bar{c}dsud+\bar{c}duds+\bar{c}dusd+\bar{c}sddu+\bar{c}sdud+2 \bar{c}sudd-\bar{c}udds-\bar{c}udsd-2 \bar{c}usdd), \\
(T_{15})^{\{23\}}_2&=&\frac{1}{8} (\bar{c}sdsu-\bar{c}dssu-\bar{c}dsus+2 \bar{c}duss+\bar{c}sdus+3 \bar{c}suds+3 \bar{c}susd-2 \bar{c}udss-3 \bar{c}usds-3 \bar{c}ussd), \\
(T_{15})^{\{23\}}_3&=&\frac{1}{8} (\bar{c}sddu-\bar{c}dsdu-\bar{c}dsud-3 \bar{c}duds-3 \bar{c}dusd+\bar{c}sdud-2 \bar{c}sudd+3 \bar{c}udds+3 \bar{c}udsd+2 \bar{c}usdd), \\
%(T_{15})^{\{31\}}_1&=&\frac{1}{8} (3 \bar{c}dssu+3 \bar{c}dsus+2 \bar{c}duss-3 \bar{c}sdsu-3 \bar{c}sdus-\bar{c}suds-\bar{c}susd-2 \bar{c}udss+\bar{c}usds+\bar{c}ussd) \\
%(T_{15})^{\{31\}}_2&=&\frac{1}{2} (\bar{c}susu+\bar{c}suus-\bar{c}ussu-\bar{c}usus) \\
%(T_{15})^{\{31\}}_3&=&\frac{1}{8} (-2 \bar{c}dsuu-3 \bar{c}dusu-3 \bar{c}duus+2 \bar{c}sduu-\bar{c}sudu-\bar{c}suud+3 \bar{c}udsu+3 \bar{c}udus+\bar{c}usdu+\bar{c}usud) \\
%(T_{15})^{\{32\}}_1&=&\frac{1}{2} (\bar{c}dsds+\bar{c}dssd-\bar{c}sdds-\bar{c}sdsd) \\
%(T_{15})^{\{32\}}_2&=&\frac{1}{8} (-\bar{c}dssu-\bar{c}dsus+2 \bar{c}duss+\bar{c}sdsu+\bar{c}sdus+3 \bar{c}suds+3 \bar{c}susd-2 \bar{c}udss-3 \bar{c}usds-3 \bar{c}ussd) \\
%(T_{15})^{\{32\}}_3&=&\frac{1}{8} (-\bar{c}dsdu-\bar{c}dsud-3 \bar{c}duds-3 \bar{c}dusd+\bar{c}sddu+\bar{c}sdud-2 \bar{c}sudd+3 \bar{c}udds+3 \bar{c}udsd+2 \bar{c}usdd) \\
(T_{15})^{\{33\}}_3&=&\frac{1}{4} (\bar{c}sdsu-\bar{c}dssu-\bar{c}dsus-2 \bar{c}duss+\bar{c}sdus-\bar{c}suds-\bar{c}susd+2 \bar{c}udss+\bar{c}usds+\bar{c}ussd),\\
(T_{15})^{\{23\}}_1&=&\frac{1}{2} (\bar{c}dsds+\bar{c}dssd-\bar{c}sdds-\bar{c}sdsd), \\
 (T''_{15})^{\{11\}}_1&=&\frac{1}{4} (\bar{c}dusu+\bar{c}duus-\bar{c}sudu-\bar{c}suud+\bar{c}udsu+\bar{c}udus-\bar{c}usdu-\bar{c}usud), \\
 (T''_{15})^{\{11\}}_2&=&\frac{1}{2} (\bar{c}suuu+\bar{c}usuu-\bar{c}uusu-\bar{c}uuus),\
 (T''_{15})^{\{11\}}_3=\frac{1}{2} (-\bar{c}duuu-\bar{c}uduu+\bar{c}uudu+\bar{c}uuud), \\
 (T''_{15})^{\{12\}}_1&=&\frac{1}{12} (2 \bar{c}ddsu+2 \bar{c}ddus-\bar{c}dsdu-\bar{c}dsud+2 \bar{c}duds+2 \bar{c}dusd-\bar{c}sddu-\bar{c}sdud-4 \bar{c}sudd\\&&+2 \bar{c}udds+2 \bar{c}udsd-4 \bar{c}usdd), \\
 (T''_{15})^{\{12\}}_2&=&\frac{1}{12} (2 \bar{c}dsuu-\bar{c}dusu-\bar{c}duus+2 \bar{c}sduu+2 \bar{c}sudu+2 \bar{c}suud-\bar{c}udsu-\bar{c}udus+2 \bar{c}usdu\\&&+2 \bar{c}usud-4 \bar{c}uuds-4 \bar{c}uusd), \\
 (T''_{15})^{\{12\}}_3&=&\frac{1}{12} (-4 \bar{c}dduu-\bar{c}dudu-\bar{c}duud-\bar{c}uddu-\bar{c}udud+8 \bar{c}uudd), \\
 (T''_{15})^{\{13\}}_1&=&\frac{1}{12} (\bar{c}dssu+\bar{c}dsus+4 \bar{c}duss+\bar{c}sdsu+\bar{c}sdus-2 \bar{c}ssdu-2 \bar{c}ssud-2 \bar{c}suds-2 \bar{c}susd\\&&+4 \bar{c}udss-2 \bar{c}usds-2 \bar{c}ussd), \\
 (T''_{15})^{\{13\}}_2&=&\frac{1}{12} (4 \bar{c}ssuu+\bar{c}susu+\bar{c}suus+\bar{c}ussu+\bar{c}usus-8 \bar{c}uuss), \\
 (T''_{15})^{\{13\}}_3&=&\frac{1}{12} (-2 \bar{c}dsuu-2 \bar{c}dusu-2 \bar{c}duus-2 \bar{c}sduu+\bar{c}sudu+\bar{c}suud-2 \bar{c}udsu-2 \bar{c}udus+\bar{c}usdu\\&&+\bar{c}usud+4 \bar{c}uuds+4 \bar{c}uusd), \\
 %(T_{15})^{\{21\}}_1&=&\frac{1}{12} (4 \bar{c}ddsu+4 \bar{c}ddus-2 \bar{c}dsdu-2 \bar{c}dsud+\bar{c}duds+\bar{c}dusd-2 \bar{c}sddu-2 \bar{c}sdud-2 \bar{c}sudd+\bar{c}udds+\bar{c}udsd-2 \bar{c}usdd) \\
 %(T_{15})^{\{21\}}_2&=&\frac{1}{12} (4 \bar{c}dsuu-2 \bar{c}dusu-2 \bar{c}duus+4 \bar{c}sduu+\bar{c}sudu+\bar{c}suud-2 \bar{c}udsu-2 \bar{c}udus+\bar{c}usdu+\bar{c}usud-2 \bar{c}uuds-2 \bar{c}uusd) \\
 %(T_{15})^{\{21\}}_3&=&\frac{1}{12} (-8 \bar{c}dduu+\bar{c}dudu+\bar{c}duud+\bar{c}uddu+\bar{c}udud+4 \bar{c}uudd) \\
 (T''_{15})^{\{22\}}_1&=&\frac{1}{2} (\bar{c}ddds+\bar{c}ddsd-\bar{c}dsdd-\bar{c}sddd),\  (T_{15})^{\{22\}}_3=\frac{1}{2} (-\bar{c}dddu-\bar{c}ddud+\bar{c}dudd+\bar{c}uddd), \\
 (T''_{15})^{\{22\}}_2&=&\frac{1}{4} (\bar{c}dsdu+\bar{c}dsud-\bar{c}duds-\bar{c}dusd+\bar{c}sddu+\bar{c}sdud-\bar{c}udds-\bar{c}udsd), \\
 (T''_{15})^{\{23\}}_1&=&\frac{1}{12} (8 \bar{c}ddss-\bar{c}dsds-\bar{c}dssd-\bar{c}sdds-\bar{c}sdsd-4 \bar{c}ssdd), \\
 (T''_{15})^{\{23\}}_2&=&\frac{1}{12} (2 \bar{c}dssu+2 \bar{c}dsus-4 \bar{c}duss+2 \bar{c}sdsu+2 \bar{c}sdus+2 \bar{c}ssdu+2 \bar{c}ssud-\bar{c}suds-\bar{c}susd\\&&-4 \bar{c}udss-\bar{c}usds-\bar{c}ussd), \\
 (T''_{15})^{\{23\}}_3&=&\frac{1}{12} (-4 \bar{c}ddsu-4 \bar{c}ddus-\bar{c}dsdu-\bar{c}dsud+2 \bar{c}duds+2 \bar{c}dusd-\bar{c}sddu-\bar{c}sdud+2 \bar{c}sudd\\&&+2 \bar{c}udds+2 \bar{c}udsd+2 \bar{c}usdd), \\
 %(T_{15})^{\{31\}}_1&=&\frac{1}{12} (2 \bar{c}dssu+2 \bar{c}dsus+2 \bar{c}duss+2 \bar{c}sdsu+2 \bar{c}sdus-4 \bar{c}ssdu-4 \bar{c}ssud-\bar{c}suds-\bar{c}susd+2 \bar{c}udss-\bar{c}usds-\bar{c}ussd) \\
 %(T_{15})^{\{31\}}_2&=&\frac{1}{12} (8 \bar{c}ssuu-\bar{c}susu-\bar{c}suus-\bar{c}ussu-\bar{c}usus-4 \bar{c}uuss) \\
 %(T_{15})^{\{31\}}_3&=&\frac{1}{12} (-4 \bar{c}dsuu-\bar{c}dusu-\bar{c}duus-4 \bar{c}sduu+2 \bar{c}sudu+2 \bar{c}suud-\bar{c}udsu-\bar{c}udus+2 \bar{c}usdu+2 \bar{c}usud+2 \bar{c}uuds+2 \bar{c}uusd) \\
 %(T_{15})^{\{32\}}_1&=&\frac{1}{12} (4 \bar{c}ddss+\bar{c}dsds+\bar{c}dssd+\bar{c}sdds+\bar{c}sdsd-8 \bar{c}ssdd) \\
 %(T_{15})^{\{32\}}_2&=&\frac{1}{12} (\bar{c}dssu+\bar{c}dsus-2 \bar{c}duss+\bar{c}sdsu+\bar{c}sdus+4 \bar{c}ssdu+4 \bar{c}ssud-2 \bar{c}suds-2 \bar{c}susd-2 \bar{c}udss-2 \bar{c}usds-2 \bar{c}ussd) \\
 %(T_{15})^{\{32\}}_3&=&\frac{1}{12} (-2 \bar{c}ddsu-2 \bar{c}ddus-2 \bar{c}dsdu-2 \bar{c}dsud+\bar{c}duds+\bar{c}dusd-2 \bar{c}sddu-2 \bar{c}sdud+4 \bar{c}sudd+\bar{c}udds+\bar{c}udsd+4 \bar{c}usdd) \\
 (T''_{15})^{\{33\}}_1&=&\frac{1}{2} (\bar{c}dsss+\bar{c}sdss-\bar{c}ssds-\bar{c}sssd),\
 (T''_{15})^{\{33\}}_2=\frac{1}{2} (\bar{c}sssu+\bar{c}ssus-\bar{c}suss-\bar{c}usss), \\
 (T''_{15})^{\{33\}}_3&=&\frac{1}{4} (-\bar{c}dssu-\bar{c}dsus-\bar{c}sdsu-\bar{c}sdus+\bar{c}suds+\bar{c}susd+\bar{c}usds+\bar{c}ussd).
\end{eqnarray*}
\end{small}
In addition, the tensor with fully symmetric index $15'$ can be derived from $6\otimes 6\to T_{15'}$.
\begin{small}
\begin{eqnarray*}
(T_{15'})^{\{1123\}}&=&\frac{1}{12} (\bar{c}dsuu+\bar{c}dusu+\bar{c}duus+\bar{c}sduu+\bar{c}sudu+\bar{c}suud+\bar{c}udsu+\bar{c}udus+\bar{c}usdu+\bar{c}usud\\&&+\bar{c}uuds+\bar{c}uusd) \\
(T_{15'})^{\{1223\}}&=&\frac{1}{12} (\bar{c}ddsu+\bar{c}ddus+\bar{c}dsdu+\bar{c}dsud+\bar{c}duds+\bar{c}dusd+\bar{c}sddu+\bar{c}sdud+\bar{c}sudd+\bar{c}udds\\&&+\bar{c}udsd+\bar{c}usdd) \\
(T_{15'})^{\{1231\}}&=&\frac{1}{12} (\bar{c}dsuu+\bar{c}dusu+\bar{c}duus+\bar{c}sduu+\bar{c}sudu+\bar{c}suud+\bar{c}udsu+\bar{c}udus+\bar{c}usdu+\bar{c}usud\\&&+\bar{c}uuds+\bar{c}uusd) \\
%(T_{15})^{\{1232\}}&=&\frac{1}{12} (\bar{c}ddsu+\bar{c}ddus+\bar{c}dsdu+\bar{c}dsud+\bar{c}duds+\bar{c}dusd+\bar{c}sddu+\bar{c}sdud+\bar{c}sudd+\bar{c}udds+\bar{c}udsd+\bar{c}usdd) \\
(T_{15'})^{\{1233\}}&=&\frac{1}{12} (\bar{c}dssu+\bar{c}dsus+\bar{c}duss+\bar{c}sdsu+\bar{c}sdus+\bar{c}ssdu+\bar{c}ssud+\bar{c}suds+\bar{c}susd+\bar{c}udss\\&&+\bar{c}usds+\bar{c}ussd) \\
(T_{15'})^{\{1112\}}&=&\frac{1}{4} (\bar{c}duuu+\bar{c}uduu+\bar{c}uudu+\bar{c}uuud),\
(T_{15'})^{\{1113\}}=\frac{1}{4} (\bar{c}suuu+\bar{c}usuu+\bar{c}uusu+\bar{c}uuus) \\
(T_{15'})^{\{1222\}}&=&\frac{1}{4} (\bar{c}dddu+\bar{c}ddud+\bar{c}dudd+\bar{c}uddd),\
(T_{15'})^{\{1333\}}=\frac{1}{4} (\bar{c}sssu+\bar{c}ssus+\bar{c}suss+\bar{c}usss) \\
(T_{15'})^{\{2223\}}&=&\frac{1}{4} (\bar{c}ddds+\bar{c}ddsd+\bar{c}dsdd+\bar{c}sddd),\
(T_{15'})^{\{2333\}}=\frac{1}{4} (\bar{c}dsss+\bar{c}sdss+\bar{c}ssds+\bar{c}sssd) \\
%(T_{15})^{\{1121\}}&=&\frac{1}{4} (\bar{c}duuu+\bar{c}uduu+\bar{c}uudu+\bar{c}uuud) \\
(T_{15'})^{\{1122\}}&=&\frac{1}{6} (\bar{c}dduu+\bar{c}dudu+\bar{c}duud+\bar{c}uddu+\bar{c}udud+\bar{c}uudd) \\
%(T_{15})^{\{1131\}}&=&\frac{1}{4} (\bar{c}suuu+\bar{c}usuu+\bar{c}uusu+\bar{c}uuus) \\
%(T_{15})^{\{1132\}}&=&\frac{1}{12} (\bar{c}dsuu+\bar{c}dusu+\bar{c}duus+\bar{c}sduu+\bar{c}sudu+\bar{c}suud+\bar{c}udsu+\bar{c}udus+\bar{c}usdu+\bar{c}usud+\bar{c}uuds+\bar{c}uusd) \\
(T_{15'})^{\{1133\}}&=&\frac{1}{6} (\bar{c}ssuu+\bar{c}susu+\bar{c}suus+\bar{c}ussu+\bar{c}usus+\bar{c}uuss) \\
(T_{15'})^{\{2233\}}&=&\frac{1}{6} (\bar{c}ddss+\bar{c}dsds+\bar{c}dssd+\bar{c}sdds+\bar{c}sdsd+\bar{c}ssdd) \\
(T_{15'})^{\{1111\}}&=&\bar{c}uuuu,\ (T_{15'})^{\{2222\}}=\bar{c}dddd,\ (T_{15'})^{\{3333\}}=\bar{c}ssss.
\end{eqnarray*}
\end{small}

\section*{Acknowledgments}

This work is supported in part by National Natural Science Foundation of China under Grant No. 12005294.

  \end{document}